**Novel results in STM, ARPES, HREELS, Nernst, neutron, Raman, and isotope substitution experiments and their relation to bosonic modes and charge inhomogeneity, from perspective of negative-$U_{eff}$ boson-fermion modelling of HTSC**

**John A Wilson**

H.H. Wills Physics Laboratory,
University of Bristol,
Tyndall Avenue,
Bristol BS8 1TL.  U.K.

**Abstract**

This paper seeks to synthesize much recent work on the HTSC materials around the latest STM results from Davis and coworkers.  The conductance diffuse scattering results in particular are used as point of entry to discuss bosonic modes, both of condensed and uncondensed form.  The bosonic mode picture is essential to understanding an ever growing range of observations within the HTSC field.  The work is expounded within the context of the negative-$U$, boson-fermion modelling long advocated by the author.  This general approach is presently seeing much theoretical development, into which I have looked to couple many of the experimental advances.  While the formal theory is not yet sufficiently detailed to cover adequately all the experimental complexities presented by the real cuprate systems, it is clear that it affords very appreciable support to the line taken.  An attempt is made throughout to say why and how it is that these events are tied so very closely to this particular set of materials.

**§1. Introduction to the STM work of Hoffman *et al*.**

Recently Hoffman *et al* [1] have uncovered in a 4K STM study of local tunnelling conductance into Bi-2212 that over the energy range 5 to 30 meV additional novel scattering information is forthcoming. Through a Fourier transformation of their real space local conductance maps they show that structured incommensurate scattering is being incurred. This scattering is quite distinct from that previously detected due to (*i*) low energy impurity effects, (*ii*) magnetic effects, (*iii*) charge stripe effects, (*iv*) vortex effects, and (*v*) the Bi-O layer superlattice modulation. The newly found modulations in the tunnelling signal are much more diffuse than that from (*v*), and the wavevectors involved are besides seen to be strongly energy dependent. The dispersion characteristics manifest in the new scattering are moreover not those of the antiferromagnetic magnons of low doping, nor likewise of the lattice phonons. The involved wavevectors ($\mathbf{q}_A$ and $\mathbf{q}_B$) are found to relate directly to the basal Fermi surface geometry, $\mathbf{k}_F$, and the scattering is coupled furthermore with tip binding energies roughly equal in value to the superconducting gap, $2\Delta(\mathbf{k}_F)$. As a result Hoffman *et al* tentatively have presented their 4K data in terms of the introduced electrons being scattered from the Bogoliubov quasiparticles. The density of states peak for the latter would then of course map out the energy below $\mathbf{k}_F$ of the bottom of the superconducting band gap as structured by the $d_{x^2-y^2}$ symmetry HTSC order parameter. However upon closer examination of the STM results we will show that whilst at every level of sample hole doping the locus from the full set of conductance modulation wavevectors indeed creates a shape in *k*-space very close to the ARPES-determined normal state Fermi surface, the associated binding energies $E_q(\theta)$ consistently are appreciably smaller than the actual gap – prior at least to the 'hot spot' being reached (see below). It will be shown that this observation is in accord with the existence of a dispersed uncondensed boson mode, as figures in the negative-*U*, two-subsystem treatment of HTSC from the current author [2]. Formal development in modelling of mixed boson-fermion and/or negative-*U* form recently has been much expanded by Micnas *et al* [3], by Domanski *et al* [4], and by Casas, de Llano, Solis *et al* [5], closer examination being made of questions relating in particular to the electronic specific heat, entropy and condensation energy data from Loram *et al* [6] and to the Uemura-type rendering of the μSR data [7].

A matter we will return to in due course but one that at this point must be registered is the misleading reading of the true value of the superconducting gap from the ARPES spectra adopted by Hoffman *et al*. In the latter spectra the leading peak should be viewed - unlike in earlier type readings of such data, and taken up now in [1] - as much more closely defining $2\Delta$ than it does $\Delta$. $\Delta$ in fact is better monitored by the mid-point of the leading edge. At optimal doping in YBCO-123 and BSCCO-2212 it is the maximal $2\Delta(\theta)$ gap values which fall just short of 40 meV. Thermally activated experimentation such as specific heat and nmr, along with phonon line-width and electronic Raman analyses [2e (§E16) and 2f (ref 36)] all concur on such a $2\Delta_o$ value ($\equiv 320$ cm$^{-1}$). This then yields the well-known strong-coupling $2\Delta_o/kT_c$ ratio of approximately 5.5. A proper reading of $2\Delta$ directs interpretation of what Hoffman *et al*'s new results mean along significantly different lines from those followed in [1]. The correct interpretation of the ARPES spectrum has been a long and complex task − one controlled by the joint effects of the local boson mode and bilayer splitting.

What is most striking about the new STM results, as indicated above, is that for any given doping level the *maximal* (*) scattering signal detected in the conductance materializes at some $E_q^*$ appreciably smaller than $2\Delta(\mathbf{k}^*,p',T=0)$ for coupled $\mathbf{q}_A^*$, $\mathbf{q}_B^*$. The latter wavevectors define a set of equivalent points within $k$-space that fall in fact well removed from the $(\pi,0)$ saddles in the band structure and indeed also from the renowned 'hot spots', sited on lines $(0,\pi)$ to $(\pi,0)$, *etc.* where those lines intersect the Fermi surface. In the spin fluctuation interpretation of HTSC the 'hot spots' are to be associated with the highly characteristic $(\pi,\pi)$ scattering physics [8], directly in evidence with the striking 'resonance mode' in inelastic neutron scattering [9]. By contrast within our own negative-$U$ interpretation the latter scattering involves not true *magnetic* spin excitation but rather a singlet-triplet spin-flip pair breaking of local pairs [2]. The maximal scattering in the new STM data, besides showing up in k-space as close to the $d_{x2-y2}$ nodes as to the hot points, does so at a binding energy $E_q^*$ that is only ≈ ½.$2\Delta_o(\mathbf{k}^*,p')$. It would appear the STM conductance signal becomes maximized where/when the dispersed boson mode of the postulated negative-$U$ model stands not too severely perturbed by either pair formation or pair dissolution, these in the model occurring in the vicinity of the hot spots and the gap nodes respectively.

These STM scattering results are employed in the present paper as point of departure against which an extended range of recent experimental results are to be viewed within the framework of the author's long-standing negative-$U$ perception of HTSC behaviour. Section 2 presents some of the earlier detailing of the model in conjunction with a full description of the STM scattering experiment and data. Comparison is made with recent ARPES results. Section 3 presents the STM scattering object as being a dispersed mode of uncondensed local pair bosons, and it contrasts these results with what has been recorded in inelastic neutron scattering events, both of spin-flip and phononic origin. Section 4 deals by contrast with the condensed boson state and with its excited plasma mode. The latter is seen as responsible for the kinking induced in the quasiparticle bands around 60 meV, and for the strong HREELS signal at this energy. The Nernst results are embraced here and again at the beginning of Section 5. The latter deals principally with the nature and consequences of the mixed valent inhomogeneity and returns to new STM work. Section 6 starts again from STM work in introducing the various roles LO phonons play in HTSC phenomena, particularly $\Sigma_1$ coupling to the plasma mode of the boson condensate. The consequences for phonon dispersion, Raman, IR, nmr, isotopic shift and high pressure results are detailed. Finally Section 7 looks more closely at the current theoretical state of play regarding the nature of HTSC in the cuprates. It includes the effect of a slight breakage of e-h symmetry in the Bogoliubov sense, and contrasts this with the strong asymmetry observed between hole and electron carrier types within the bank of known superconductors. Shell-filling effects have been contrasted with the band Jahn-Teller type as regards the sourcing of exotic non-BCS type superconductivity. Indication is given of how spin-fluctuation, phonon and Marginal Fermi Liquid modelling of HTSC each fall short of what may be accomplished by pursuing the boson-fermion, resonant negative-$U$, two-subsystem approach.

## §2. Relationship of the new STM data with 2Δ(θ) gap data from ARPES.

Hoffman *et al* observe that the new diffuse peaks appearing in the Fourier analysis of the STM conductance data clearly look to reflect the geometry of the Fermi surface. In fact the two wavevectors **q** detailing the scattering effects in *k*-space seem in elastic fashion to span the Fermi surface between equivalent points in the zone, with $\mathbf{q}_A$ parallel to the Cu-O bonds and $\mathbf{q}_B$ set at 45° to these. The $\mathbf{q}_A$ thus span across the occupied arms of the Fermi sea while the $\mathbf{q}_B$ span between these arms across the unoccupied parts of the zone. Accordingly note that the $\mathbf{q}_A$ lie at 45° to the Bi-O superlattice modulation, the sharp diffraction spots from which are very visible in the STM conductance F.T.'s [1; fig 3]. The above assignment is in full accord with the observation made in [1] that as the (hole) doping is augmented so wavevector $\mathbf{q}_A$ decreases while the partnering $\mathbf{q}_B$ increases. Plotted in figures 4a,b of [1] is just how the measured scattering amplitudes change with the wavevector magnitudes $|\mathbf{q}_A|$ and $|\mathbf{q}_B|$, and how the scattering peak positions progressively are shifted in *k* space as the binding energy of the STM tip is incrementally augmented. By combining these two plots one is able, upon accepting the above dispositions for $\mathbf{q}_A$ and $\mathbf{q}_B$, to trace out the Fermi surface. The outcome was not actually displayed in [1] and is given now in fig. 1. Included in the figure is a circle about the zone corner which for an idealized, homogeneous, band-like situation in the 2D limit would correspond to a hole doping $p_h$ of (1)+ 0.16, the band filling to yield maximum $T_c$ in all cuprate HTSC systems.

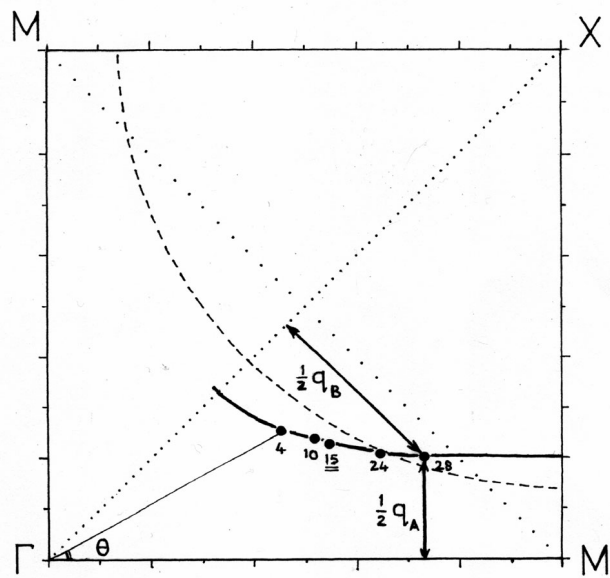

*Figure 1.*  Shown in a quadrant of the Brillouin zone is the locus defined by the coupled elastic scattering wavevectors $\mathbf{q}_A$ and $\mathbf{q}_B$ (see text) extracted from Fourier analysis of the STM conductance real space mapping secured at 4 K by Hoffman et al [1] for the case of slightly underdoped ($T_c = 78$ K) $Bi_2Sr_2CaCu_2O_{8+\delta}$. Also shown is the hypothetical circular 2D Fermi surface that would correspond to a hole count relative to half filling (the dashed line) of 0.16, as for optimaly doped HTSC material. The *experimental locus obtained here is very similar to that generated in LDA band structural work for the bilayer structured material at optimal doping, as it relates to the fuller of the two sheets to derive from the two Cu-O planes per unit cell (i.e. the bonding interlayer combination). The 'hot spots' are where the M-M dotted line intersects the Fermi surface on the band structural saddles near the M points ('$\pi$,0', etc.), and the superconducting $d_{x2-y2}$ node lies on the line from 0,0 to '$\pi$,$\pi$', close here to one third of the way from zone centre to zone corner. Angular $\theta$ values from the $k_x$-axis read off from this plot for each data point are subsequently used in constructing figure 2.*

Of course the HTSC compounds are not ideal metals but very highly correlated, are not homogeneous but chemically mixed-valent and dynamically charge- and spin-striped [2e,d], and are not in the 2D limit but hold measurable 3D character. In the Bi-2212 case the latter automatically will involve direct intra-bilayer interaction and for a standard material this is going to lead to a two-sheeted $d_{x2-y2}$ symmetry Cu-O $pd\sigma^*$ band complex. Here the fuller component (of *electrons*, and the one supplying the outermost electron F.S. sheet) will be the one for which the intra-bilayer interaction entails *c*-axis 'bonding' as opposed to 'antibonding' phasing between the two $pd\sigma^*$ wavefunctions that issue from the pair of Cu-O chessboard arrays per unit cell. Following much initial searching for this 3D splitting several more recent ARPES works employing better resolution are now able to detect both components to the $pd\sigma^*$ band [10-12], at least for optimally and overdoped material. Since the same groups however when working with Bi-2223 still observe only two dominant spectral features, not three [13], and yet more tellingly with Bi-2201 continue to report two clear features, not just one [14], this justifies our continued adherence to the position recently re-expressed by Campuzano, Norman and Randeria [15] in their very comprehensive review of the ARPES work; namely that the universal peak and hump circumstance apparent in all these spectra is the outcome of strong correlation and scattering, and comes about independently of any multisheet coupling. The consensus to emerge from the more refined ARPES determinations now as regards the overall shape of the Fermi surface is that the latter departs appreciably from the circular form of the idealized surface present in figure 1 towards a cross, retracted somewhat in the vicinity of the 45° nodal directions and correspondingly inflated in the saddle point regions. It is pleasing to observe then that this is precisely the kind of modification upheld in figure 1 by the new STM data. The curve seen there is very comparable in form to that for the fuller F.S. component of the LDA band structure calculations [16], and again to that traced by the uppermost spectral feature (sharp peak) in the ARPES work. In the ARPES data the two key spectral features ('peak and hump') are separated at the saddle points by ~ 90 - 140 meV, this dependent upon doping level. LDA band structural work [16] produces a bonding/antibonding splitting at the saddles under a coherent bilayer-type interaction of up around 300 meV, but in contrast the new ARPES work indicates a much smaller value in fact of about 85 meV. By symmetry this value becomes steadily reduced to zero on moving to the 45° directions. It is for very different reasons that the $\pi$,$\pi$ direction is as well the orientation taken on by the superconducting nodes [2b].

At this juncture it is well to recall that de Haas-van Alphen-type resonances have not yet been reported in the HTSC cuprates, even for that most favourable of cases YBa$_2$Cu$_4$O$_8$ (Y-124) [17], and in all probability this will not be achievable either for very highly overdoped material, seeing the parallel null result gained for the 3D *s*-electron superconducting mixed-valent system (Ba/K)BiO$_3$ [18]. All HTSC materials are characterized by chronic scattering and this is not just because they are poorly formed or intrinsically nonstoichiometric. With PrBa$_2$Cu$_4$O$_8$, upon greatly compromising there the potential within the Cu-O basal *planes* for metallic conductivity (and indeed superconductivity), this chain-bearing material becomes rendered a notably better metal, more coherent and Fermi-liquid-like, even in its *c*-axis direction [19]. Recall with Pr-124 that despite it macroscopically being crystallographically perfect, one witnesses the basal plane electric and magnetic behaviour to be totally transformed under the strong Pr*4f*-Cu*3d* hybridization. The latter inhibits production of the RVB spin-singlet condition so vital to allowing HTSC to arise.

Within the author's negative-*U* two-subsystem perception of HTSC [2] the very severe S = 0 pair scattering experienced in HTSC *Y*BCO-124, -123, etc. by the basal carriers near the saddle points and hot spots constitutes the very means to gaining local pair formation in the key shell-filling negative-*U* process. Saddle point scattering is the origin not just to the superconductivity but in large measure too to the highly anomalous 'normal' state properties, in particular to the *p*-type sign for the carriers reinforcing the strong spin and charge pseudogapping, and likewise to the $T^{-2}$ dependence of their mobility to very high temperatures [2e, 20]. As has been elaborated upon recently by Hussey [21], the HTSC cuprate systems exhibit large basal scattering anisotropies bearing enormous *T*-independent prefactors. Moreover customary poor-metal resistance saturation no longer is pre-eminent. The above chronic saddle-point scattering is demonstrated to add in parallel to a more normal component in evidence once away from the saddles. As it happens, given the particular geometries of crystal structure and Fermi surface, even the best of carriers, namely those running in the nodal directions, are susceptible to strong *e-e* scattering over into the saddle-point sinks, as was portrayed in fig 3 of [2b]. It is not surprising therefore that the HTSC materials retain significant local character through to such high doping levels, despite standard LDA band structure analysis ascribing to the $d_{x2-y2}$ *pd*σ* band a bare width in excess of 4 eV. Such width reflects the large degree of *p*/*d* mixing being incurred as the copper *d*-states drop down towards closure through the oxygen based *p*-states [22]. The above extremely strong *e-e* scattering persists in fact to well out beyond $p_h$ = 0.3 [23], beyond where the novel superconductivity can any longer be sustained.

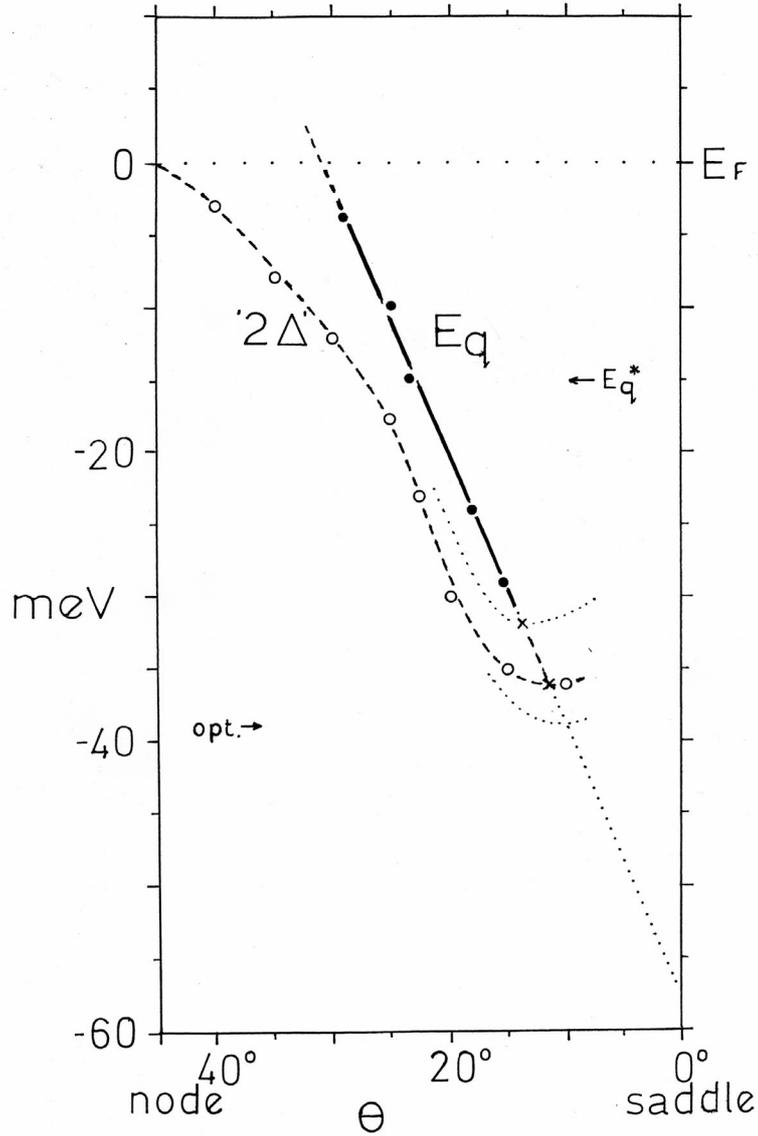

*Figure 2.* *The angular variation within the Brillouin zone octant from the nodal to the saddle directions is plotted for the binding energies $E_q$ (at 4 K) corresponding to the conductance data points located at coupled wavevectors $q_A$ and $q_B$ in figure 1. The more or less linear dispersion presented is that for a somewhat underdoped sample. The line appears to shift slowly upwards with a rise in sample hole doping content. The diffuse STM scattering signal is steadily lost as on the one hand the experimental conditions approach the Fermi energy and on the other hand are pushed back down towards the 'hot spots' on the saddles. In the present case the sharpest scattering peak signal was registered for the intermediate condition $\theta \approx 23°$. The above angular behaviour is to be contrasted with that included in the figure of the first sharp peak in the ARPES data under comparable conditions, this constructed from the data of Mesot et al [24]. The latter curved plot (dashed) is taken to be a moderately close guide to the functional form of the superconducting gap $2\Delta(\theta)$ – not $\Delta(\theta)$ as stated in [24]. It is not justified, moreover, to take directly the non-sinusoidal form of this plot as representing faithfully any true deviation from a strictly $d_{x^2-y^2}$ form for the superconducting order parameter.*

In figure 1 the tip energy has been included as running parameter; also indicated is the point at which the STM signal maximizes. The latter arises well away both from the saddle points and the 45° nodal directions. Actually the present scattering falls below detection considerably in advance of reaching either such limiting position. In figure 2 the next step is made of displaying the variation of the above $E_q$ as a function of $\theta$ around the Fermi surface, $\theta$ being measured about the zone centre anticlockwise from the $\pi,0$ saddle-point direction. This plot permits comparison to be obtained between energies $E_q(\theta)$ and $2\Delta(\theta)$, the latter being set approximately by the first sharp peak in the ARPES spectra formed below $T_c$. Angular peak position plots have been produced for a variety of under-, optimally and over-doped BSCCO-2212 samples by Mesot et al [24]. It is such information which we now incorporate in figure 2, although reading this here as much more closely related to $2\Delta$ than to $\Delta$ (as originally taken in [24]). Both the $E_q(\theta)$ and the $2\Delta(\theta)$ results are quite difficult to quantify precisely and that calls for a certain amount of smoothing of the raw data. Nonetheless several features stand out from the plots in figure 2: (*i*) $E_q(\theta)$ effectively is a linear function, (*ii*) $2\Delta(\theta)$ as emphasized by Mesot et al [24] is (as extracted) far from being pure $d_{x2-y2}$ in form, (*iii*) despite $2\Delta(\theta)$ bowing towards reduced $|2\Delta|$ versus an idealized sin $2\theta$ form, it ultimately does turn toward zero as $\theta \to 45°$, (*iv*) $E_q(\theta)$ however goes to zero much in advance of $\theta = 45°$, (*v*) the maximum value of $2\Delta(\theta)$ does look to be realized in the vicinity of the hot spot where the Fermi surface crosses the $(0,\pi)$-$(\pi,0)$ tie line near $\theta = 12°$. There exists no Fermi surface of course close to the $\theta = 0°$ saddle-point direction, in particular in the 2D limit. What now is viewed as highly significant is that across the central range of $\theta$ the STM data points $E_q(\theta)$ clearly reside, for common $\theta$, at binding energies appreciably reduced as compared with $2\Delta(\theta)$, identity being gained only at the stage of maximal $2\Delta(\theta)$ reached with the hot spot location [2b]. All such hot spot locations appertaining to the various sample underdopings seem to line up along the extrapolated $E_q(\theta)$ linear plot. Optimal doping is noted to occur much in advance of this extrapolation driving the hot spot point right back to the saddle axis $\theta = 0°$. A rough examination reveals that the latter circumstance would not be reached until $p \approx 0.3$. In actual HTSC systems such a doping level happens in fact to be where the last signs of superconductivity are to be recorded. Within the present modelling that occurs because by $p \approx 0.3$ there no longer remains any well-defined two-subsystem (mixed-valent) character: the materials are transformed to more standard Fermi liquid behaviour, and the high correlation negative-$U$ HTSC scenario has been rendered inoperative [2].

**§3. The local pair uncondensed boson mode as scattering object.**

Norman and coworkers have for some time advocated that the peculiar form of the ARPES spectrum below $T_c$, with its peak, dip, hump structure, points to interaction of the quasiparticle states with some bosonic mode [25]. The natural implication of the presence of such a feature is that it relates critically to the superconductivity if not actually causing it. Not only does the above spectral activity figure at the gap antinodes, but at the gap nodes too there is evidence of a perturbation of simple behaviour, a clear kink being evident in the quasiparticle dispersion curves and growing more marked below $T_c$. When Johnson et al first reported this behaviour they implicated spin fluctuations [26]. Subsequently it was suggested by Lanzara and coworkers [27], from analogy with what is found in a standard strong-coupling superconductor like Mo [28], that phonons were responsible. The energy now however is too large (~ 60 meV) to be appropriate for phonons in general, and the situation would have to be more akin to some specific optic phonon coupling, as is present in a CDW/PSD system like 2H-TaSe$_2$ [29]. Although this action might relate in the cuprates to incipient stripe formation, it is not then evident why the kinking would grow sharply at $T_c$, to display a component rising in magnitude as the superconducting order parameter. Despite spin-singlet pseudogapping being strongly in evidence [9], Norman and colleagues appear to have settled for a magnetic attribution to the mode's origin [30,8a], largely in view of its similarity in energy to the much discussed 'magnetic resonance' excited at **Q** = ($\pi,\pi$) in neutron inelastic spin scattering − for optimally-doped Bi-2212 a quite sharp feature at around 43 meV [9]. The alternative perspective already advocated in [2a,b] is that the mode being sensed in the ARPES measurements is associated with local bosonic pairs - pairs created through the shell closure, charge fluctuation, negative-$U$ process. The latter proceeds in line with the energetics laid out in [2h,22], which it is claimed in [2c] are substantiated by the laser pump-probe experiments [31] and the thermomodulation spectroscopy [32] from Stevens *et al* and Holcomb *et al* respectively. This interpretation subsequently received further support from similar, more wide-ranging optical work by Little *et al* [33], by Kabanov, Demsar *et al* [34], and above all by Li *et al* [35], results examined at length in [2a]. The situation throughout has been perceived as a fragmented two-subsystem one (see fig 4 in [2h]), often at low doping being driven dynamically towards stripes [2e,d]. This highly perturbed geometrical condition (not entirely unlike that in a quasicrystal), when taken in conjunction with a negative-$U$ pair resonance stationed in close degeneracy with the Fermi energy, assures that the level of particle-particle scattering will be extreme. A high flux of interchange between fermionic and bosonic states, as well as between boson states that are condensed and boson states which remain outside the Bose condensate, is inevitable.

Calculations of a negative-$U$ Hubbard type consistently have indicated that when, as here, the overall *effective* negative-$U$ value (per pair) emerges as being of the order of the bandwidth [36], the greatest $T_c$ values will at that stage be encountered. In the present case |$U_{eff}$| is indeed like $W(d_{x2-y2}) \approx 3$ eV [2a-c]. As was listed in [3,4,5], theoretical (homogeneous) negative-$U$ boson-fermion equilibrium models recently have been much extended beyond the original work of Micnas, Ranninger and Robaszkiewicz [37] and of Friedberg and Lee [38]. A key matter specifically to be addressed now by de Llano, Solis and coworkers [5b] is that of boson pairs possessing non-zero center-of-momentum along with individual bosons instantaneously remaining unincorporated into the Bose condensation. The latter pairs, unlike the condensate itself, may of course show considerable dispersion under the electronic/thermal activation. A sizeable population of uncondensed bosons is able to develop in step with the number of condensed bosons (both negative-$U$ pairs and seeded Cooper pairs). Such uncondensed bosons were suggested earlier in [2a] as being likely source of the extra component of a.c. conductivity observed to show itself in roughly order parameter form below $T_c$ within the 100 GHz time domain transmission spectroscopy of Corson *et al* [39]. As the system becomes more underdoped the *fractional* population of uncondensed pairs grows as the negative-$U$ state slips to higher binding energy. This is seen in the growth in the $2\Delta^*$ value recorded as a pseudogap.

In developing their bosonic mode modelling of the ARPES data a considerable advance recently has been made by Eschrig and Norman [30b] in incorporating properly into the self-energy analysis of the mode/quasiparticle interaction the effects of the multi-layer coupling. There emerges a clear picture of just how the peak, dip, hump structure develops within both the bonding and antibonding components to the ARPES signal. It is the uppermost (i.e. antibonding) component which it is shown bears the dominant peak marking fairly closely the maximal value of $2\Delta(\theta)$ – just 32 meV in the underdoped ($T_c$ = 65K) BSCCO-2212 sample employed in [30b]. With optimally doped material the comparable value has risen to 37 meV. The latter binding energy falls still some way short though note of the 43 meV of the 'resonance mode' which so characterizes the polarized inelastic neutron spin-flip scattering [9a]. Earlier I have associated that particular resonance with complete destruction of a local pair boson. Such identification would match the above numbers, with 37 meV per pair being allotted to the separation of the pair of electrons into individual fermions, and with the additional 6 meV demanded for the imparted spin-flip to one partner, in accord with the minimum spin gap magnitude. The individual components of the pair find themselves back then near the hot spots on the two saddles from which originally they were drawn in the negative-$U$ pair production process. The above pair elimination calls for a net momentum transfer to its electrons of $(\pi,\pi)$; $(\pi - \varepsilon, \varepsilon)$ to the one emerging onto the *x*-axis saddle and the balance of $(\varepsilon, \pi - \varepsilon)$ to the one coming onto the *y*-axis saddle (see [2b, fig 3]). For the case of bilayered YBCO and BSCCO what particularly is noteworthy in the neutron scattering process is that it predominantly occurs accompanied by a large incommensurate *z*-axis momentum transfer, corresponding in direct space to the Cu-O bilayer spacing. Eschrig and Norman [30b] following the lead of Fong, Bourges *et al* [9] interpret the latter detail as evidence of a dominant magnetic aspect to conditions in the material, and regard the bonding and antibonding channels in these bilayer systems as being associated with 'even' and 'odd' parity magnetic spin coupling respectively. They pursue this direction however in spite of the sizeable spin gap, and with no evidence in optimally or overdoped material for any genuinely magnetic correlation length divergence towards low *T* [40]. In contrast the negative-$U$ point of view would see the *z*-axis momentum transfer as being demanded by charge symmetry as a local pair is undone and its components separated into nearest-neighbour cells. The two electrons with being left spin-parallel by the action of the neutron are no longer allowed to occupy the same Cu-O coordination unit (as $d^{10}p^6$). The process is illustrated in detail in figure 3.

It needs to be pointed out here that the above *z*-axis 'complication' is not in any way essential to the viability of the HTSC mechanism, it relating solely to pair break up. It simply is a symmetry requirement imposed by the crystal structure. When as with Tl-2201 one has a single layer structure the neutron spin-flip scattering peak centres on the $k_z$ = 0 plane [9b]. It remains to be observed how this structure will simplify also the details of the ARPES spectrum for Tl-2201.

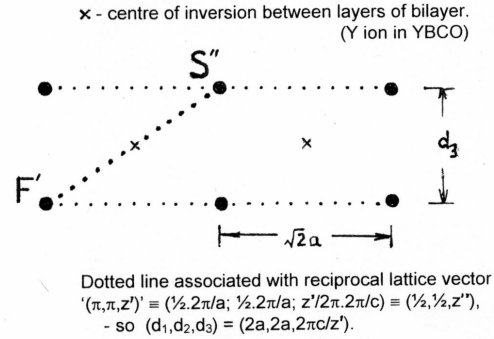

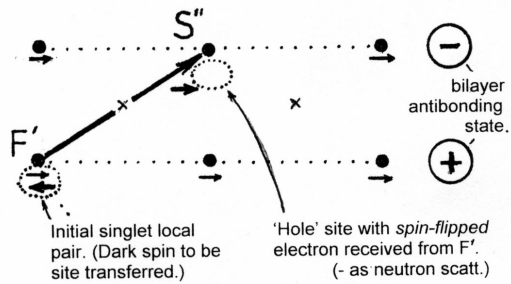

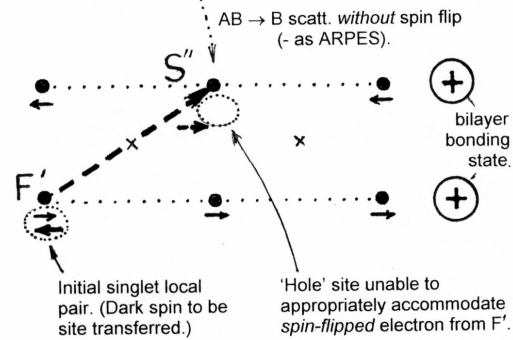

*Figure 3.* The symmetry conditions governing the way in which the bosonic pairs decompose under inelastic neutron spin-flip scattering within the bilayer structures of $YBa_2Cu_3O_7$ and $Bi_2Sr_2CaCu_2O_8$. The situation is shown in vertical (110) section. The appropriate process picks up the $k_z$ component $z' = 2\pi c/d_3$, where $d_3$ is the spacing between Cu-O planes in the bilayer, and it occurs in the antibonding bilayer charge phasing channel.

The {π,π} point composite bosons are able by virtue of their negative-$U$ standing to take on an energy that is near-degenerate with the Fermi energy. Through elastic boson-boson collisions these bosons are from this state capable of being rendered **k** = 0 entities of like energy, and as such are open in turn to becoming seeding agents for Cooper pair formation from further Fermi surface quasiparticles in a more standard **+k**/**-k** fashion. The addition of such Cooper pairs will drive up somewhat the system-wide superconductive coherence length. In the above way the negative-$U$ bosons potentially can introduce an overall level of pairing wherein the triggering zone-edge component may well constitute numerically the minority species.

Besides the two types of condensed composite negative-$U$ boson, the derivative Cooper pairs, and any remaining unpaired fermions (in particular present for our d-wave circumstance), one must as well anticipate, as stated earlier, heavy undisrupted bosons that have been excited from their condensate through phonon interaction, clearly also in play in HTSC processes. Inelastic neutron scattering experiments bring to light in fact strong modification to the dispersion of certain basal longitudinal optic phonons at very short wavelength ($|\mathbf{k}| \geq \frac{1}{2}(0,\pi)$) [41]. Such phonons possess the momentum necessary to transfer a composite boson into the saddle regions for the (closely degenerate) Fermi surface as a thermally/electronically excited bosonic quasiparticle outside either the **k** = 0 or π,π condensates. Since being of appreciable lifetime these excited bosons are going to hold binding energies somewhat less than for the Bogoliubovons at the same locations in *k*-space. Wherever such dispersed excited bosons either would drop below the chemical potential in the superconducting state (as near the saddles) or rise above the Fermi energy for the unpaired quasiparticles (as near the nodal *k,k* axis) the excited composite particles will there become unstable and disintegrate. In the above manner we perceive then how it is possible to see materialize a bosonic mode having many of the attributes displayed by the novel STM scattering results introduced in §2. The detected mode departs upwards from the 2Δ(θ) gap energies in an ever increasing fashion as one shifts away from the hot spot location toward larger θ. While the condensate coupling holds to a finite 2Δ(θ) binding right round to the nodes, the excited bosons disintegrate considerably before reaching the θ = 45° orientation. The excited boson mode stands sharpest for mid-range θ values because there the excited state lifetime is longest, least affected by the high electronic activity proceeding around the nodes and saddles.

**§4. The condensed boson mode.**

It has at this point to be emphasized that the above dispersed mode is distinct from the bosonic mode concentrated upon by Eschrig and Norman in their ARPES-related work [8a,30]. The latter mode runs virtually undispersed throughout the outer part of the zone. In the present work that mode will be associated directly with the condensate itself of local pair bosons and lies immediately below the dispersed mode of *un*condensed pairs. The extrapolation of the line relating to the dispersed mode in figure 2 down into the (π,0) saddle-point region serves to place the ground state 'bosonic resonance' (labelled $\Omega_{res}$ by Eschrig and Norman) as approaching 60 meV; i.e. at a somewhat greater binding than the maximal superconductive gap energy per pair $2\Delta_o$ for this *under*doped sample. Conversely for the *over*doped ARPES sample dealt with in [8a],

Eschrig and Norman's self-energy analysis indicates the little dispersed mode in that case to sit just *above* the slightly diminished $2\Delta_o$ level operative there. The more underdoped the system is the further below the Fermi level the local pair negative-$U$ state resides. We shall return shortly to this second mode and to its relation to the neutron scattering resonance peak.

With somewhat underdoped 2212-BSCCO it has been detected that 60 meV in fact is an energy echoed in the self-energy driven 'kink' feature readily visible in the ARPES-determined quasiparticle dispersion curves once clear of the saddle direction [26]. There under the prevailing crystalline and condensate symmetries the superconductive and bilayer gapping effects become reduced and the state structure simpler. The persistence of this observed band kinking to well above $T_c$ speaks of the continued presence of minority negative-$U$ local pairs there, while the loss at $T_c$ of the Meissner effect, *etc.* reflects the extinction with that point of the majority Cooper pair population, and along with it a relinquishing of global phase coherence. In the present view the spin pseudogap existing to 100 degrees and more above $T_c$ is to be ascribed not so much to superconductive inter-pair phase fluctuation as to dynamic RVB spin gapping. As is stated in [2], this RVB gapping is perceived as the means to preparing the system toward the production of local-pair spin-singlets, in addition to being key to their preservation. An absence of any anomalous peaking in the c-axis IR conductance just above the superconducting energy gap within underdoped, yet 'well-formed', Y-124 would support that the fluctuational behaviour manifestly present far beyond $T_c$ is not primarily superconductive in nature [42]. Full participation of the bulk of fermions in the seeded Cooper pairing is required before the systems can take up an overall condition rendered somewhat more standard in its transport properties. Once below $T_c$ the dramatic fall away in the chronic saddle-point scattering of local pair creation and excitation brings about very rapid growth in the residual electronic mean free path, this disclosed for example in the thermal Hall data [43] or in the very steep drop off in nmr relaxation rate [44].

Exactly what the nature of the low dispersion $\Omega_{res}$ mode is and of its relationship to the neutron scattering peak has been most hotly debated [45]. The uniqueness of both features within the superconducting field reveals the very close links to the local pair condensate itself. This likewise is true as regards the strong and very specific coupling into the problem of the basal-plane Cu-O bond-stretching longitudinal optic phonons. Over outer parts of the zone these particular phonons show themselves severely depressed in energy [41,46], and in a fashion quite dissimilar to that for some standard Fermi surface dictated soft-mode CDW/PLD circumstance [47] (see §6). The fact that the effect (as with the kink in the quasiparticle dispersion above) is observed as well in the *k,k,*0 direction rules out any connection with stripe phase formation. The virtually dispersionless nature of the ARPES-revealed mode as contrasted with the mode detected in the STM work is striking. The former brings to mind the phenomenon of second sound in liquid $^4$He. Upon looking into the relevant literature it soon is uncovered that comparable excitations have long been postulated for strong-coupling superconductors. What is more, their study actually was extended some years ago to 2D geometry by Belkhir and Randeria [48] in the wake of developments with the cuprates – and in fact was performed within a negative-$U$ context.

What has been explored in reference [48] is how the collective mode spectrum of a (homogeneous) superconducting system can evolve as one progresses from BCS weak coupling to the hard core boson régime across the intermediary crossover régime, wherein the cuprates clearly reside. The paper would indicate the existence of a smooth linkage between the Anderson mode established in weakly coupled BCS superconductors and the Bogoliubov sound mode applicable to the extreme local pair limit. Belkhir and Randeria adopt a site-homogeneous negative-$U$ Hamiltonian and then go forward within a generalized RPA formalism to examine the crossover régime. They turn specifically to a 2D geometry and make also the important incorporation of screened and unscreened Coulombic interactions, highly relevant to short-range quasiparticle pairing. The outcome is that the collective mode excitations in the crossover region exhibit small but finite dispersion as $\sqrt{q}$. The energy of the mode, while being raised somewhat above the hard-core boson plasma excitation (which would be rather low-lying being governed by $1/m_b \propto t^2/U$), displays very considerable departure from the high-lying Goldstone-like zero-dispersion behaviour of the Anderson mode, in consequence of the increased mass of the bound pairs within the strong-coupling lattice-type modelling. For conditions where $T_c$ maximizes (i.e. where $U/W \approx 1$ [36, 3]) the mode is able to acquire a dispersion which amounts to a not insignificant fraction of $W$. Simultaneously it is moved down in absolute energy to well below the Fermi velocity-dictated plasma energy of the Anderson mode for weak coupling. Under the real *in*homogeneously doped circumstances of the HTSC cuprates one can envisage the above plasma frequency as becoming depressed not just because of the raised value of the effective mass for the individual bosons but additionally by virtue of their relatively low and spatially variable population density.

In probing the dielectric/optical response of a material it is well known a plasma excitation is best monitored through the electron energy loss (EELS) function $\mathrm{Im}\{-1/(1+\varepsilon(\omega,\mathbf{q}))\}$. It is this same function that one looks to too for information regarding the longitudinal optic phonons. HREELS work on a metal demands high surface quality, in effect it monitoring the surface resistivity. The low temperature cleavage of BSCCO affords one excellent opportunity to obtain reliable plasma excitation results, a project in fact carried through back in 1992 by Li, Huang and Lieber [49]. Sure enough what was reported is a single strong spectral peak centred at 60 meV, this displaying moreover a thermal behaviour which clearly associates it with the superconductivity. The above workers actually related the signal to pair breaking. Notwithstanding this they demonstrated the 60 meV peak to possess a $T$ dependence which despite seeming to collapse sharply at $T_c$ would if treated as a mean-field BCS feature in fact extrapolate to an 'onset' temperature of 150 K. Accordingly it seems that our boson mode engenders appreciable energy loss only as it becomes heavily damped upon the engagement at $T_c$ of the majority Cooper pair population.

The presence of a multi-sourced pseudogapping up to temperatures of 150 K and beyond is endorsed in a great many types of experiment, not least optical, but the cleanest indicator to date that this temperature range holds some 'superconductive' content is supplied in the very recent Nernst effect measurements from Wang *et al* [50]. The Nernst effect involves the production of an electric cross-potential (y-axis) signal as the passage of charge down a (x-axis) thermal gradient is subjected to a large and mutually perpendicular (z-axis) magnetic field – the thermal equivalent of the Hall effect. The signal issues from normal carriers in addition to entities associated with the superconductivity, but of course whenever the latter exist they are dominant. The above paper actually focusses on the signal which may arise from flux vortices once these enter their 'liquid state', depinned from the lattice and able to drift in the thermal gradient. Because the signal Wang *et al* record clearly extends up 20 or more degrees above $T_c$ it is apparent however that a local pair boson attribution becomes there the more appropriate.

### §5. Expressions of the mixed-valent inhomogeneity.

What additionally is so instructive about these Nernst data [50] is that they supply a direct means of assessing $H_{c2}$ (at which applied field the strong Nernst signal component naturally is taken to zero). Although with most HTSC samples $H_{c2}$ lies well beyond the 30 teslas experimentally available in [50], it is discovered, as a result of there emerging simple scaling rules for both reduced fields and temperatures, that the zero temperature critical field can be extracted as a function of (under-)doping right up to 150 tesla (for $p = 0.08$). The $H_{c2}(0)$ values so deduced are directly convertible to the corresponding coherence lengths $\xi_o$ { $= (\phi_o/2\pi H_{c2})^{½}$} and manifest the latter as a monotonically rising function of doping $p$ (i.e. of metallicity), $\xi_o$ being close to 20 Å (or $5a_o$) at optimal doping but only 10 Å down at $p \sim 0.05$. This reflects that as the participation of Cooper pairs is advanced the extreme local character of the superconductivity becomes somewhat relaxed. Despite $n_s$ growing in proportion to $p$ through this doping range [7], the pairing force itself ultimately peaks, and hence accordingly so does $T_c$. For some time it has been established from the analysis of the electronic specific heat following Loram and coworkers [6] that the overall condensation energy per dopant charge is diminished to either side of a 'critical doping' (just above the optimal for $T_c$) and in particular towards the underdoped side. As was pointed out in [2h], concentrations of hole doping of the parent Mott insulator very close to the optimal level are facilitated in this optimization of properties by virtue of the percolation limit they represent within the two-subsystem basal plane geometry.

In view of the above it is rather strange Loram, Tallon and Liang in a recent preprint [51] have chosen to play down the fragmented circumstances prevailing in the cuprate systems, particularly when underdoped. They declare that the rather sharp character to what occurs at critical doping would imply a more uniform state, and they advance their view by reference to sharp signals seen in nmr work, specifically for [81]Y and [17]O. However the type of disorder being addressed in [51] is restricted to that associated solely with static doping disorder and frozen charge segregation, as at first sight the recent STM results may appear to flag [52]. In all HTSC systems (bar $YBa_2Cu_4O_8$ and fully oxygen loaded $YBa_2Cu_3O_7$) there inevitably exists the substitutional or interstitial disorder introduced to lift the carrier dopant count to the desired level. At low temperatures this atomic substituent disorder *is* frozen and generally random. However the key question in regard to the metallic and superconducting properties (Y-124 included) is what is the spatial distribution of the coupled hole content per Cu beyond unity. What is registered becomes then a matter of time scale for the particular experimental probe employed. Wherever charge carrier dwell times are long compared with the characteristic probe time one will obtain a broad and multi-sited signal, whereas, and notably with nmr, if local dwell times are (relatively) short some motional narrowing of the signal ensues. It has been demonstrated in recent $^{63}$Cu zero-field nmr work (from LSCO) [53] just how the spectra gathered upon thorough analysis clearly uphold a situation in which fluctuating regions (if not domain walls) of segregated charge indeed arise. It would be most valuable to have this work repeated now for high quality Hg-1201, a system in which the hole doping is secured via interstitial oxygen rather than cationic substitution. The mercury materials, as with all others examined, certainly present evidence of a $^1/_8$ anomaly [54]. Within the fluctuational mêlée the *local* effective doping level has proved in the LSCO case to be such that it can stand far different from the mean value [53], to such a degree that magnetic moments can in fact emerge at select sites even within mildly underdoped material. This is what it has been argued in [2d] leads to the weak spin-flip effects recorded in μSR [55] and neutron diffraction [56]. Indeed it is the Swiss cheese view of the situation, supported at one time by Loram, Tallon and coworkers [57], although given here a dynamic time scale – more like grilling cheese. The percolation threshold is a sharp threshold and one clearly of considerable import for global superconductivity within these strongly coupled systems. The μSR results early on showed how the averaged $n_s$ count achieved under hole doping $p$ climbs linearly, and with $n_s$ also $T_c$, as sensed via transport related properties – in the case of μSR via the screening controlled penetration depth [7]. The specific heat work [6] showed under thorough analysis how, nonetheless, a fully proportionate condensation energy in advance of critical doping is not attained; only with the latter is proper superconductive coherence between the different entities and structural fragments in the system fully attained. It is in this way that HTSC systems come to display glassy characteristics below $T_c$, especially when underdoped [58].

    A striking close up affirmation of the local conditions prevailing in the superconducting state of HTSC materials is to be found in the low temperature, high spatial and energy resolution, STM mappings of BSCCO samples given in a fairly recent paper from Davis and coworkers [59]. Let us inspect this remarkable new data from the perspective of the present paper. Spatially one observes a condition highly fragmented as regards energy gap magnitude, this within a sample of bulk crystallographic near-perfection. The latter categorization neglects (*i*) the dopant excess oxygen, (*ii*) some Bi on Sr sites, and (*iii*) the Bi-O layer supermodulation. These atomic 'imperfections' are all frozen in at 4.2 K and their electrostatic potential creates a time-invariant backdrop for the dopant charge. The latter adjusts and fluctuates at the unit cell level to establish the dynamic two-subsystem environment in which the boson-fermion negative-*U* superconductive action proceeds. In the energy-resolved tunnelling one finds some nano-regions to settle into gaps of around 39 meV with others up around 58 meV, and a bridge of values between the two. In [59] the former locations are termed α-regions and the latter high gap regions β-regions. From what we have argued already we see that these regions are to be associated with Cooper pair and with local pair predominance respectively. In line with this identification Davis and coworkers report the following: (*i*) the β-regions diminish in relative *areal* weighting as hole doping progresses

− in a lightly overdoped sample they form just 10% of the field of view as compared with 50% in an underdoped sample of $T_c$ = 79 K; (*ii*) the conductance within the majority carrier, small gap α-regions is appreciably greater than in the local pair β-regions, (*iii*) the local conductance signal amplitude and the coupled maximal gap size accordingly convert in *anti*phase upon STM tip passage from the one type of region to the other; (iv) only the low gap Cooper pair α-regions display an impurity (≤ ½% Ni) intragap resonance signal under a *positive* applied voltage (+18 meV), such as would come with Bogoliubovon-like hole state behaviour. While these results in general are supportive of our present modelling there is one aspect that by their nature they cannot show − namely any appreciable organization away from random towards stripe geometry, due to the transient form of the latter organization under charge hopping, evident in the Cu nmr work [53]. The intermediate gaps nonetheless provide some mark of this activity. The α and β regions are remarkably sharp-edged however and, what is more, are remarkably consistent in their interior gap magnitudes, in this justifying our 'two-subsystem' terminology.

The above discussion of the mixed-valent inhomogeneity and its consequences for these systems is very much in line with 'simple' theoretical modelling attempted already within the context of a negative-*U* Hubbard Hamiltonian. Working with *s*-wave coupling geometry and employing the Bogoliubov-de Gennes procedure, Suvasini, Gyorffy and coworkers obtained interesting initial results [60] and the problem subsequently has been carried through in some detail for intermediate coupling (although neglecting Coulomb interactions) by Ghosal, Randeria and Trivedi [61]. One of the latters' main results is that for the *s*-wave case an effective global order parameter persists, and this in conjunction moreover with a non-vanishing spectral gap. It clearly would be of much value now to know conversely if for strong local *d*-wave coupling, a vanishing or near-vanishing spectral gap would persist, as experiments on the cuprates seem largely to suggest. Contrasting work within the *t-J* model with *d*-wave coupling, but adhering to the spinon-holon view of HTSC, also has made investigation of the consequences for the system of the introduction of strong local disorder [62] (in that paper parametrized through the (screened) effective distance of the dopant ions from the active Cu-O plane).

**§6. A role for phonons.**

Just published this month by Davis and coworkers is yet another STS paper with novel results bearing strongly upon the above [63]. Following on from the Hoffman paper [1], they now include a reconstruction of the F.S. such as was given earlier in figure 1, but secured this time by identifying and simultaneously fitting *all* the symmetry equivalent elastic F.S. scattering vectors within the zone, $q_1$ to $q_8$, and not just the shortest two detected before (then labelled $q_A$ and $q_B$). This new more detailed work continues however to assert an identity between the quasiparticle Bogoliubovon states and the STM conductance registered states. The latter in contrast we will continue to claim actually lie at somewhat smaller binding energy than the former (assessed through ARPES): we still feel these states in reality do not display the dispersion and intensity characteristics to match Davis and colleagues' interpretation of events. The experimental mode dispersion curve constructed in [63] remains much like that in figure 2 above, and if anything it shows a slightly counter-sigmoid form, not giving one any confidence at all that the mode will swing round to zero binding by $\theta = 45°$. Perhaps the best argument to support the Bogoliubovon interpretation offerred in [63] by McElroy *et al* is that quite similar results now are revealed to show up at *positive* bias also. We return to this matter in due course.

Following on from the discussion of previous sections it is best to deal first however with the observations which have been appended to the new paper [63] in its closing paragraph. The authors put on record there, along with figure 4, that the direct space conductance mapping signal, *g*(**r**,ω), under specific local conditions exhibits a novel fine-grained 'tweed' structure. The period of this tweed is just *one* unit cell, and it materializes within the map only when and where the tip tunnelling energy equals the locally relevant superconducting gap *maximum*. Both italicized observations serve to indicate that the tweed structure is associated with strong scattering activity proceeding at the zone edge. Scattering under a reciprocal lattice vector G signals the occurrence of a coupled umklapp phonon process, and one that arises seemingly regardless of whatever the *local* gap size maximum at the saddles might actually be. In the present model such lattice involvement already has been mentioned in regard to satisfying momentum conservation as local pairs (*i*) are shifted between the (π,π) negative-*U* states and the saddles, and again (*ii*) convert at the saddles into +**k**/-**k** Cooper pairs. In each case the net momentum per pair required for these changes is close to (1,0) or (0,1).

Just which phonons might be involved here is to be gleaned from inspection of the phonon dispersion curves obtained via inelastic neutron scattering. As already has been noted, there is found to occur appreciable phonon softening (relative to the corresponding Mott insulating material) of certain longitudinal optic modes towards the basal zone edge [41]. Moreover noticeable intensity changes are found to develop between branches and also within a branch as a function of temperature. More striking still, there would at first sight seem to be present 'extra' features in the spectrum beyond the branches anticipated from a standard phonon normal mode analysis for the given compound. Extensive neutron scattering data clarifying these matters have just been published by Chung et al [64], and these deserve close examination now with express regard to phonon coupling to the local pair boson modes discussed above.

Chung *et al*'s work [64] actually deals with $YBa_2Cu_3O_{6.95}$. Because very large crystals are required for the neutron work, the samples of the orthorhombic material were not *a,b* detwinned. The endemic twinning has the unfortunate effect of superimposing the TO and LO phonon branches sensed by the neutrons ($\mathbf{I} \propto (\mathbf{Q}_n\cdot\mathbf{\epsilon})^2$) for the given propagation and displacement directions, and hence mode separation within the experimental data demands careful scrutiny.

**Table 1:** *Characteristics of zone boundary phonons in YBa$_2$Cu$_3$O$_7$ propagating in basal plane perpendicular and parallel to chains; i.e. along a axis (irreducible representations Σ) and b axis (irreducible representations Δ) respectively. Information drawn from Chung et al [64], figs. 1 and 4 and text. (Note small errors exist in figure 1 there, with 1e being labelled LO rather than TO, and superfluous oxygen atoms appearing in 1c, 1e and 1f at the chain level).*

| figure label [64] | in plane propag$^n$ vector (re chain) | displacements (re propg$^n$ vector : LO vs. TO) | displacements or 'polarization' (re chain) | dominant vibration type (re Cu-O bonds) | phonon mode irreducible representation | Γ pt. energy (meV) |
|---|---|---|---|---|---|---|
| 1a | x (⊥) | x (// : LO) | x (⊥) [Raman B$_{2g}$] | (⊥ chain) planar stretching | Σ$_{1,4}$ | 72 |
| 1b | y (//) | y (// : LO) | y (//) [Raman B$_{3g}$] | (// chain) planar & chain stretching | Δ$_{1,4}$ | 66 |
| 1c' | y (//) | x (⊥ : TO) | x (⊥) | (⊥ chain) planar stretching | Δ$_{2,3}$ | 72 |
| c'' | x (⊥) | y (// : TO) | y (//) | (// chain) planar stretching | Σ$_{2,3}$ | 66 |
| 1d' | y (//) | y (// : LO) | y (//) [Raman inactive] | (// chain) chain stretching | Δ | * |
| d'' | x (⊥) | y (⊥ : TO) | y (//) [Raman inactive] | (// chain) chain stretching | Σ | * |
| 1e' | x (⊥) | z (⊥ : TO) | z (⊥) [Raman active] | (⊥ chain) apical stretching | Σ$_1$,Δ$_1$ | 62.5 |
| e'' | y (//) | z (⊥ : TO) | z (⊥) [IR active] | (⊥ chain) apical stretching | Σ$_4$,Δ$_4$ | 69 |
| 1f' | x (⊥) | x (// : LO) | x (⊥) | (⊥ chain) bending of both chain & plane. | Σ | 43 |
| f'' | y (//) | x (// : TO) | x (⊥) | Ditto | Δ | * |

LO : longitudinal optic, *i.e.* displacement vector // propagation vector, and Cu and O displacements in bonds in antiphase.

Σ vs Δ splitting is from presence of chains destroying tetragonal symmetry.

The Σ representation wavevectors are //*a* (*i.e.* are ⊥ to chains), while the Δ are //*b* (*i.e.* are // to chains).

Σ$_1$ and Σ$_4$ relate to there being two basal planes per unit cell. (Δ$_2$ and Δ$_3$ likewise),

- of which only ungerade (antisymmetric) inter-layer combination seen generally for integer $l$.

Basal modes associated with phonon propagation down the crystallographic *a* axis take irreducible representations designated by Σ, while those running in the *b* axis (chain) direction carry the label Δ. The highest energy axial branches $^{LO}$Σ$_{1,4}$ and $^{TO}$Δ$_{2,3}$ issue from a zone centre Γ state of point group symmetry *B*$_{2g}$ at 72 meV, which Raman work also detects. Modes $^{LO}$Δ$_{1,4}$ and

$^{TO}\Sigma_{2,3}$ emerge from a $\Gamma$ state of symmetry $B_{3g}$ lying at 66 meV, which again Raman spectroscopy detects. Already this 6 meV splitting, a consequence of the presence of the chains, is quite sizeable, and effects become more marked once well away from the zone centre. The subscripts on $\Sigma$ and $\Delta$ above specify states covered by particular irreducible representations within the axially relevant subgroups, and in each case a couple of phonon states are to be distinguished, the outcome of there in Y-123 being two $CuO_2$ planes per unit cell. Because inter-layer coupling is in general weak in the HTSC structures, such pairs of states see their degeneracy lifted noticeably only as there develops further more interesting electronic coupling into one or other partner. In order to keep track of the situation close study of table 1 is recommended. The key action is witnessed in the longitudinal modes, as befits charge coupling, and is especially marked for one of the above $^{LO}\Sigma_{1/4}$ branches. These particular $^{LO}\Sigma$ branches each have both their propagation and their polarization vectors aligned perpendicular to the chains (*i.e.* parallel to *a*), while the two complementary $^{LO}\Delta_{1,4}$ modes conversely have both vectors parallel to the chains (*i.e.* parallel to *b*). No odd behaviour is evident in the associated TO branches, whether in the form of any softening or splitting or intensity change as a function of temperature, and the same looks to be true for the bond-bending modes at lower energy. At first sight the apical modes seem equally uninteresting, but note here that there is present a nearly constant offset right across the zone of 6-9 meV between the Raman and I.R. active partners $\Sigma_1$ and $\Sigma_4$ (or $\Delta_1$ and $\Delta_4$), with the simpler symmetry $\Gamma_1$-compatible states taking the lower energy. Note that these apical modes in spite of formally being transverse are dominantly bond-stretching in nature like $^{LO}\Sigma_{1,4}$ and $^{LO}\Delta_{1,4}$. The latter two sets of LO branches each exhibit considerable downward dispersion into the body of the zone, but what really marks out their behaviour as being unusual is their intensity change with wavevector, portrayed in [64] figure 5. As against a standard shell model fitting to the phonon data, there is found to occur experimentally an enormous gain in intensity of certain branches in the vicinity of the hot spots near the zone edge, together with energies shifted down to ~ 55-57 meV (see [64] figures 5 and 6). While the $^{LO}\Sigma$ and $^{LO}\Delta$ branches each move into this range, the differing manner of their doing so is most revealing. What is apparent is that the degeneracy between the $\Sigma_1/\Sigma_4$ interlayer combinations suddenly becomes lifted halfway across the zone, and it remains so right through to $\pi,0$. The outer segment of the $\Sigma_1$ branch looks to have dropped in energy below the $^{LO}\Delta_{1,4}$ branches, which possibly remain unsplit. (There of course is no interaction here between the $\Sigma$ and $\Delta$ branches - they are in mutually perpendicular orientations within each twin domain.) The means whereby the depressed segment of the $\Sigma_1$ branch is strongly acquiring intensity clearly is some further $\Sigma$ mode residing in the vicinity of 55 meV – and this is not the bond-bending phonon mode, which stays little changed in intensity right across the zone ([64], fig. 5b). The changes wrought in the intensity of the anomalous $\Sigma_1$ phonon branch segment possess two very revealing characteristics; (*i*) as known for some time [41] they become more pronounced with the hole doping, *p*, (*ii*) they grow with temperature reduction below $T_c$, actually tracking there the general form of the superconducting order parameter. Yet more telling, however, is the sudden onset to this strong interaction as the phonon wavelength drops below $5a_o$ or 20 Å. (N.B. in the

colour figures 2,7 and 9 in [64], red denotes an (ω,k) sector showing gain in intensity upon *cooling* down through $T_c$, whilst blue denotes an intensity loss.)

It becomes necessary to examine then what is the nature of the $\Sigma_1$ mode to which the bond-stretching $^{LO}\Sigma_1$ phonon branch couples so strongly. That there additionally is a clear although less dramatic coupling to $^{LO}\Delta_1$ and to the apical bond-stretching phonons $\Sigma_1/\Delta_1$ would implicate direct charge coupling. Our discussions of §3 immediately point to the plasma-like mode of the local pair boson condensate. The latter mode looked to reside, remember, at binding energies sited just less than for the bosonic local pair condensate ground state, and was deemed to be responsible for the kinking introduced near $\mathbf{k} = \mathbf{k}_F$ at binding energy ≈ 55 meV in the ARPES extracted quasi-particle dispersion curves – a feature becoming more marked below $T_c$. This energy value for the plasma mode near $\mathbf{k} = \mathbf{k}_F$ is in keeping with the energy of the (π,π)-imparting, resonant spin-flipping excitation by neutrons of individual pairs, bringing about their ejection from the $k = 0$ superconductive condensate and ultimately a return of their components to $E_F$ at the saddles – an energy in YBCO$_7$ of 41 meV per pair. Such numbers would imply that the pair plasmon mode actually disperses slightly downwards (*i.e.* to bigger |ω|) from the zone centre along $k_x$ and $k_y$ towards the zone edge within these d-wave superconductors. (Such behaviour would parallel that of the 41 meV spin resonance excitation itself, which as was discussed at some length in [65] disperses to numerically smaller energies upon a decrease in the crystal momentum input being sustained.) The fact that effects comparable to the above occur as well in the phonon dispersion behaviour of LSCO [41] means they do *not* in YBCO$_7$ enter solely as a consequence of the crystallographic presence of the chains. The standard shell model does not generate such a strong divergence as that observed to develop between and among the Σ and Δ modes involved, whether LO or TO.

Why the $^{LO}\Sigma_1$ phonon mode with its displacement vectors perpendicular to the chains should couple more strongly with the plasmon mode than does $^{LO}\Delta_1$, with its chain-paralleling displacements, would seem largely to be a matter of compatible symmetries. The former mode involves, note, the bumping together of the chains, and this effects carrier passage between the charge stripes which in YBCO follow the chain direction [66]. More significant is the reason why the really strong coupling should set in only as $k$ passes beyond ~ ½(π/$a_o$), *i.e.* as λ drops below 4$a_o$. The latter distance is, recall, expressly equal to the pair coherence length ξ for optimally doped HTSC cuprates. As pointed out in [48], it only is for λ values inside this limit that the local pair plasmon mode is able properly to form and accordingly then to couple with the lattice.

Very short wavelength charge pair oscillations are not confined to the above basal plane states. In the bi-, tri- and higher order layer HTSC compounds there is below $T_c$ at least partially coherent superconductivity in the *c*-axis direction, with pair transfer between adjoining CuO$_2$ layers ≈ 4 Å apart. These cation occupied spaces are without oxygen and are somewhat more ionic than the covalent CuO$_2$ planes, providing lower local dielectric constants and damping. For tri-layer systems and above, Munzar and Cardona [67] very recently have drawn attention to the fact that certain 'out-of-phase' interplane charge oscillations become Raman active (see their figure 1).

Pointedly the appropriate phonon modes are below $T_c$ seen to acquire very large intensity and often very substantial frequency modifications are incurred upon passage through $T_c$ (*e.g.* for Hg-1234 the $A_{1g}$ + $B_{2g}$ (*x',x*)-polarized phonon excitation peak at 390 cm$^{-1}$ plummets by 50 cm$^{-1}$). Munzar and Cardona demonstrate, what is more, that the effect of these interplane plasma oscillations is able to produce a Raman scattering efficiency of sufficient magnitude as clearly to be origin to the anomalous form of the $A_{1g}$ electronic Raman feature, so long to have plagued interpretation of the latter type of Raman work [68].

Comparable effects to the above have been identified too in *c*-axis I.R. results [69], where once below 150 K a broad peaking is observed to develop in the c-axis conductivity, centred about 50 meV and with the neighbouring phonon peaks simultaneously falling in intensity. Note through all the myriad means of investigating the small excitation energy range up to 1000 cm$^{-1}$ (120 meV), careful distinction must be made between those effects associated with the spin pseudogapping of the DOS (which is not particularly temperature dependent, and is sensed for example in nmr Knight shift experiments) and those lower energy effects around 400 cm$^{-1}$ (or 50meV) which develop not too far above $T_c$ and are associated with a sharp reduction in the chronic basal scattering to afflict the fermionic quasiparticles (and are sensed for example in nmr relaxation rate experiments). Both types of electronic change scale with $T_c(p)$ [70], although for very different reasons. The former change relates largely to pair breaking or rather its quenching under RVB creation, and the latter change relates to pair formation commissioned by the hybrid growth in coupling between fermions and negative-*U* bosons.

Because the pair plasma mode is interacting so directly with the lattice vibrations, and strongly once below $T_c$, it is not surprising that oxygen isotope effects have been extensively recorded for HTSC systems [71]. However the latter effects are quite different in form from those coming with the standard, phonon mediated, retarded interaction of the BCS mechanism. Indeed it is observed under *p* change within a given HTSC system that the isotopic shift in regard to $T_c$ itself pointedly is dropping to zero at the composition to yield the highest $T_c$ – or, more likely, the highest condensation energy per pair and shortest coherence length, the critical doping $p \approx 0.19$. Very significantly, however, there is in operation at precisely this stage a strong isotopic mass exponent below $T_c$ as regards the penetration depth $\lambda$ ($\propto n_s/m^*_b$). Because $n_s$ is not appreciably affected by isotopic substitution this means an isotopic shift must be operative deriving from the bosonic effective mass $m^*_b$. Plainly the *electronic* masses are responding here, via the plasma mode coupling discussed above, to the eigen-energy changes imposed upon the phonon modes. On cooling from 300 to 4 K, $m^*_e$ itself has been evaluated to mount steadily from 2 to 4 $m_e$ [69].

Since the kinking in the dispersion curves, and likewise the development of the peak, dip, hump structuring to the ARPES energy spectrum, each become more pronounced with increase in layering number, *n*, across a sequence like Bi-2201, Bi-2212, Bi-2223 [72], one would much like to discover what happens to the above penetration depth isotope effect. The most probable scenario is that these layering effects actually issue from the known reduction with *n* in the basal plane lattice parameter, this rendering the negative-*U*, closed-shell state more established, yet without here any rise in pair breaking, as would be introduced through underdoping. It correspondingly is recorded that uniaxial pressure applied *within* the basal plane produces a steep growth in $T_c$ [73], and therefore potentially directly in $n_s$.

As is to be expected from the current modelling, the mass enhancements arising in the saddle regions will be greater than those encountered in the nodal regions of the Fermi surface. This plus sharp growth in renormalization of the imaginary part of the self energy below $T_c$ under the action of the boson mode is seen very clearly in the new ARPES data (scanning from π,0 across $\mathbf{k}_F$ towards π,π) just published by Kim *et al* [74]. Additionally apparent in this data is that no fundamental difference of response exists between the bonding and antibonding interlayer coupled sub-bands. The kinking is more apparent in the former simply because it is the fuller. Above $T_c$ the antinodal coupling constant rapidly is seen to become much reduced (though still around 1) and the mass (but not the scattering rate) then is more or less isotropic. Note the quasiparticle mass renormalization resulting from the self-consistent action of the electron pair mode manifests quite different thermal characteristics to what would result if the mode were primarily phononic in nature, or one of spin fluctuations.

**§7. More on the nature of the HTSC mechanism.**

In the present form of modelling what in large measure dictates the sensitivity of response to parameter changes like pressure and doping is the precise location of the double-loading negative-*U* fluctuational state with respect to the Fermi energy – or once below $T_c$ to the chemical potential. Remember the small energy ~ -50 meV by which the state is to sit below $E_F$ is but a tiny fraction of $|U_{eff}|$, which in the present case is ≅ 3 eV (per pair) [2a,c]. In turn the latter is less than half the actual state adjustment energy accompanying double-loading shell closure, since about 4 eV is negated in the requirement to 'back off' the Coulomb repulsion energy. Accordingly a high sensitivity of $T_c$ to all parameter changes is to be expected if indeed these remarkable effects establish themselves as one reaches close degeneracy between the negative-*U* state and $E_F$, with its ready interconversion between charge fermions and bosons.

This degenerate condition very recently has been referred to by Domanski [75] as a 'Feshbach resonance'. It is demonstrated formally in [75] that when the boson and fermion subsystems are disentangled, via application of a continuous canonical transformation, the residual fermion-fermion quasiparticle interaction itself presents highly resonant structure above $T_c$, which for $T < T_c$ becomes diminished in magnitude to that of the superconducting gap. A further key result for our purposes obtained in [75] is that the overall boson-fermion system foregoes the particle-hole symmetry which characterizes the standard Bogoliubov analysis of the simpler BCS mechanism.

Domanski is not the only one to investigate this symmetry breakage between the hole and electron Bogoliubov related pair states of the superconducting phase, the matter being extensively examined too by Batle *et al* in reference [5a]. One consequence of the shifting away in the Boson-Fermion model from the *e-h* symmetric BCS limit and the mean field approximation, introduced by allowing bosonic pairings to exist and propagate independently embedded in the Fermi sea, is that uncondensed boson modes occur, and these possess a novel dispersion character, in that for the 2D case their energy varies linearly with **K**, the centre of mass momentum for the pairs. (When Cooper pairs move in vacuum, as in the simple BCS treatment, their dispersion is quadratic). The velocity for this uncondensed mode of bosonic pairs is comparable in size to the Fermi velocity of the unpaired quasiparticles. It is this linearly dispersed mode of uncondensed pairs that it has been claimed in §1-3 of the present paper is that being sensed in the STM conductance scattering experiments from Hoffman *et al* [1] and now McElroy *et al* [63]. Because the 2D Fermi wavevector $k_F$ is virtually circular about $(\pi,\pi)$, a plane will intersect the related binding energy cylinder in approximately a straight line, just as the experimental results appearing in our figure 2 (or figure 3c of [63]) determine. The above mode is of appreciably greater velocity than for the plasma mode for the condensed bosons addressed subsequently in §§4-6. The velocity of the uncondensed mode being proportional actually to the pair coupling strength is automatically quite substantial in the HTSC materials. Because the pairing in these d-wave systems is dominated by action at the extensive saddles in the band structure, the interaction parameter $\lambda = N(E_F).V$ becomes considerable even if $V_{eff}$ itself is not enormous. Indeed Batle *et al* find that $T_c$ values of 100 K may be attained for $\lambda$ values of only ~ ½ upon employing energy-shell coupling ranges comparable to $\varsigma\omega_D$ (*i.e.* ~ 50 meV with $E_F$ ~ 1 eV). The authors of ref. [5a] as a result loop their argument around in favour of phononically driven coupling, in line with the orientation adopted by Lanzara *et al* in [27]. In that case however there would be no compelling reason to depart from the Bogoliubov *e-h* symmetry of the traditional mechanism.

   Batle *et al* [5a] cloud the situation somewhat further by pointing out, after Hirsch [76], that the great majority of known superconductors possess *p*-type Hall coefficients, as against most non-superconducting metals being *n*-type in their transport properties. This carrier electron-hole asymmetry, however, is far from synonymous with the BCS related usage of those terms. The present author already has commented on how the preponderance of *p*-type superconducting materials is likely to be associated with chemical bonding effects, even within the simplest of compounds. Where such effects become very evident is in the high pressure, semimetallic, homopolar bonded forms of elements like B, Si, P, S and I [77], and still more recently Li [78]. One might include here cluster structures like the $C_{60}$ [79] and $Ga_{84}$ [80] derivatives and the graphite-sulphur composites [81]. The most striking example of bonding-driven pairing outside the closed-shell effect of the cuprates and bismuthates is the case of $MgB_2$, where the holes in the σ-bonding B-B band clearly are responsible. As with the cuprates, the fact that the lattice responds to the pairing interaction – here in the unique reaction of the $E_{2g}$ bond-breathing mode – does not mean one should invert the terminology and revert to an electron-phonon ascription. Those interested further in $MgB_2$ are directed to the work of Pietronero and colleagues [82] and of Mazin and Antropov [83], dealing with these band Jahn-Teller type effects and the non-adiabatic, anharmonic lattice coupling. Note when talking about 'chemically' driven superconductivity one ought to take care to distinguish between the above homopolar bonded systems (of which β-HfNCl organo-solvated by $Li^+$ ($T_c$ = 25½ K) constitutes a further recent case [84]), and those reliant upon closed shells and disorder-forestalled disproportionation, such as the cuprates and bismuthates. An additional very recent example of the latter type would look to be $Na_xCoO_{2+\delta}$ (reported $T_c$ of 31

K [85]). This latter instance would revolve around the high stability of the low-spin $t_{2g}^6$ closed subshell configuration, as discussed in § 7.2.9/10 of [2h'].

Let us now return to the question of *e-h* symmetry in the Bogoliubov sense - or, rather, the lack of it. The breaking of such symmetry is a feature of the Boson-Fermion modelling, and one might question whether it finds clear cut experimental expression anywhere. Actually what has long been known is that tunnelling signals in the HTSC systems are not symmetric in positive and negative bias, unlike for standard BCS superconductors. Domanski and Ranninger indeed have addressed just such matters regarding tunnelling in one of a series of recent papers to employ the B-F model [4c]. Despite formal criticism of their theoretical procedures by Alexandrov [86], the general level of progress encountered in this approach suggests that the technical problems pointed to in [86], of singular divergences and cancellations, probably can be removed by making the modelling somewhat more complex, say through incorporating the action of the spatial inhomogeneities of doping and stripes present in the real systems. The presence of uncondensed as well as condensed bosons would appear well-established from the gigahertz a.c. conductivity measurements of Corson *et al* [39;see 2a], and these are a feature that Tan and Levin [87] now have taken forward formally to account not only for the Corson results, but those also of Wang *et al* [50] concerning the Nernst effect in BSCCO, as has been suggested above in §4. Formally they show how up to temperatures considerably higher than $T_c$ a mimicking of the Kosterlitz-Thouless behaviour in the former case and persistence of the Nernst signal in the latter case strongly uphold the existence there of local pairings. As indicated already, what fluctuates here, in underdoped systems in particular, is principally the superconductive phase angle between poorly communicating tight pairs, though not for quite the reasons proposed by Emery and coworkers [88]. In this connection Tan and Levin prove that the transverse thermoelectric coefficient relevant in the Nernst effect is a much more sensitive probe of the local electronic pairing activity than say are the diamagnetic fluctuations. The latter remain not dissimilar in degree to the Aslamazov-Larkin behaviour within a standard superconductor. Ranninger and Tripodi very recently have extended the theoretical modelling of the B-F scenario in this direction [89], now permitting a degree of itineracy to the hard-core bosons: they then track the decay of coupling between the subsystems as a function of underdoping, and the shift away there from amplitude to phase fluctuations.

If the above discussed excited bosonic 2*e* pair modes exist for all $T < T^\dagger$ (where the fluctuation crossover temperature $T^\dagger$ is, note, not synonymous either with the RVB spin gap temperature or the charge pseudogap temperature), there ought to be some positive indication of the corresponding independent 2*h* modes. It is our belief that the McElroy STM paper [63] in fact could contain just such an observation, a matter we flagged for subsequent comment at the end of the first paragraph in §6. McElroy *et al* state in [63] that their + and - 14 meV scattering results are 'identical' (see figure 2H vs. 2F). Because of the diffuse nature of those results it would appear as yet not feasible to confirm the as-anticipated difference between the positive and negative bias results without appreciably more care being taken. Nonetheless the general level of difference found between the ordinary $\pm V$ tunnelling results is encouraging. It would now be extremely interesting to try to follow the STM scattering signals to higher temperature - if that is manageable without surface contamination.

One thing very clear through all this work is that the reason the HTSC problem has been with us so long is that it is a truly complex problem. The previously simplest of experiments repeatedly debouche into a remarkable pan of intricacy, which it may or may not be profitable to pursue. The nmr results were an early marker of this [53,90] and the ARPES data more recently have witnessed a similar transition from being treated as beguilingly straightforward to being recognized as highly complicated in nature [91]. What may be said, though, as with the evolution of this paper, is that, wherever one probes, the boson-fermion, two-subsystem, negative-$U$ approach appears able to cope with more problems than it generates. It is sufficiently complicated of itself to hold the required degree of flexibility to pursue all the various twists and turns which the experimental work exposes. It is self-evident that neither the phonon scenario of HTSC nor the spin-fluctuation one possess the necessary degree of built-in complexity to do this. I would refer those still thinking along the latter lines to examine the 'inversion' of the photoemission spectra just presented by Verga, Knigavko and Marsiglio [92] as they attempt to account for the observed kinking in the quasiparticle dispersion curves. The spectral form of $\alpha^2.F(\omega)$ for the mode coupling derived there within an Eliashberg type treatment (see figure 6) is quite unlike that expected for spin fluctuations, and in addition it contains far too much weight at high energies to be relevant to phonons alone.

One of the chief reasons for seeing the spin-fluctuation scenario actively supported for so long is its perceived compliance with neutron scattering data. Wherever the latter has been dominated however by data coming from LSCO (in consequence of the large crystals available), that is unfortunate since LSCO is the most ionic of all HTSC systems, resulting from its particular counter-ions [2f]. The RVB spin gap is there particularly small ($\approx$ 10 meV), and the spin gap temperature actually falls somewhat below the likewise small $T_c$ [2b]. This has meant many of the observations made on cooling LSCO down below $T_c$ have been regarded as appertaining to the superconductivity when in reality they should have been attributed to the RVB spin gap. The gap studied by Lake *et al* in [93] is, note, dispersionless. Additionally at helium temperatures the spin scattering peaks which they register at $E > \Delta_{sp}$ develop there to define more sharply a 'spin coherence length' which is independent of **q** and close to 32 Å or $8a_o$. The latter is of course just about the size of the domain in the stripe structure being settled into with the establishment of the spin gap, as depicted in [2d, fig 1]. The observed cluster of four incommensurate inelastic scattering 'satellites' comes to strongly decorate the $k$-space point $(\pi,\pi)$ in consequence of the RVB 4-spin plaquet correlations. It was emphasized in [2d] that the approximately $8a_o$ size of the incipient domain structure is not Fermi surface governed but is controlled simply by the numerology of the doping charge concentration.

Millis and Drew [94] recently have drawn attention to the seeming incompatibility between the self energy broadened MDC peaks observed in the new high resolution photoemission work and the optically determined conductivity. The latter conductivity is considerably better than what the former widths in customary circumstances would match, a dichotomy not expected if events were to be covered by the spin fluctuation scenario. However, as we have seen, the strong resonant (i.e. elastic) scattering activity operative on the $k_F$ saddles associated with fermion-to-boson inter-conversion will not contribute towards restricting the optical conductivity. The large quasi-elastic scattering term has earlier been commented upon by Abrahams and Varma [95]. They deemed that out-of-plane 'impurities' had to be responsible - presumably the dopant ions. However our view of events is more 'intrinsic' yet than this. Besides the above sizeable elastic (and strongly angular dependent) term there of course remains the very substantial inelastic scattering activity between the carriers and between the carriers and the lattice. These effects go forward to produce those strong and very characteristic resistivity terms - linear both in $T$ and $\omega$ [96] - behaviour which Varma early on referred to in his Marginal Fermi Liquid formulation of HTSC behaviour as issuing from a 'magic polarizability' [97]. The carrier scattering is at least an order of magnitude more intense than what would occur if the transport properties were simply as covered by a standard Boltzmann treatment with the given LDA band structure. It is not just the unusual $T$ and $\omega$ dependencies to the observed transport properties which are in question, but, above all, the sourcing of their strikingly large prefactors. The situation for the d.c. conductivity has long been recognized, but recently Varma and Abrahams [98] have offered a formal treatment within the framework of MFL of how both the Hall and magneto-resistance signals likewise are able to emerge as being swamped by the novel scattering. They have attempted through a small angle scattering treatment to generate the well known HTSC results $\cot\theta_H$ ($\propto \mu_H^{-1}$) $\propto T^2$ and $\Delta\rho_B \propto T^{-4}$, but this as they now acknowledge is not the correct way to reach these relations [99]. As was stressed in [2e] the latter are not standard Fermiology results; indeed they constitute clear expression of that intense fermion-boson activity which precipitates HTSC within the present resonant negative-$U$ scenario. The two-particle, two-subsystem, umklapp and DOS pseudogap aspects to the problem clearly are going to have to be incorporated explicitly.

Let us now review the present theoretical situation. The magic polarizability looked to in the MFL theory, with its rather momentum independent and structureless form, energetically as thermally, is in accord with a resonant interconversion/quantum fluctuation as present in the negative-$U$ scenario. The action we have seen extends to the lattice, including the accommodation of momentum conservation as quasiparticles move between the saddles, the zone corners and the zone centre. The appropriate zone edge phonons are observed to couple into the various charge processes. The local pair bosons are still present to some degree above $T_c$ and form dispersed propagating modes within the pseudogap. Below $T_c$ the local bosons do not pass completely into the **K** = 0 condensate, and a pair of excited dispersed modes for 2$e$ and 2$h$ Cooper-like pairings persist, in addition to the excited sound-wave like excitation of the condensate itself. The condensate holds many more pairs than the dopant count, as more standard Cooper pairing around the Fermi surface is induced, although the effective overall pair coherence length is kept very small (with $H_{c2}$ high). Because of the boson-fermion degeneracy and interconversion, the condensate interacts very strongly with the quasi-particle dispersion curves to produce there kinks in all directions somewhat below $E_F$. The pairing action itself proceeds by virtual excitation from the band structural saddles to the top of the band at the zone corner, a state when empty of high energy under bonding correlations. The doubly loaded state however is greatly relaxed in negative-$U$ fashion by virtue of its association with band closure and a complete restructuring locally of all the valence band energy states. The latter comes in response to the termination of antibonding effectiveness as regards the principal Cu-O binding – the essential chemistry in the problem [2b,2h]. It is a matter of chance that for these particular square-planar cuprate materials the $^{10}Cu_{III}{}^{2-}$ double-loading state happens to fall into close resonance with $E_F$. This appears not so for the case of the 3D metallic mixed-valent system $La_4BaCu_5O_{13+\delta}$. The crystal structure of the HTSC materials is especially favourable in that it produces a strong zone edge saddle point very close to $E_F$. From the latter it is easy then to draw off considerable numbers of bosonic pairings of the approximate type $e(0,\pi) + e(\pi,0) \to b(\pi,\pi)$, and this without any participation of phonons or spin fluctuations. Phonon participation occurs as bosons return to the saddles under the negative-$U$ 'relaxation', or subsequently pass from the saddles over into the $K = 0$ condensate. The phonons involved in these charge coupled processes are longitudinal optic in nature. Similarly the bosonic mode excitations as plasma oscillations involve in the LO branches, and accordingly become registered via $\mathrm{Im}(-1/\varepsilon)$ in the HREELS experiment.

The 40 meV 'resonance peak' recorded in the neutron scattering experiments [9] is not a bosonic mode capable of supplying the 'glue' for HTSC. Indeed it marks the disruptive excitation of S = 0 superconducting pairs, involving S = 1 spin flipping, as a pair is taken back from $K = (0,0)$, via $(\pi,\pi)$, to its component fermions near $(\pi,0)$ and $(0,\pi)$. The observed $z$-axis component to this excitation in YBCO-123 and BSCCO-2212 is not to do with magnetic spin coupling but with appropriate accommodation of the two fermionic spins within the bilayer unit cell. The paper by Abanov, Chubukov and colleagues [45b] which in preprint form was entitled 'What the $(\pi,\pi)$ resonance peak can do' [100], acknowledges the spin-exciton view of events expressed there to be quite distinct from that of antiferromagnetic magnon mediation. Nonetheless they continue still to look towards the action of Fermi sea spanning to build up the response function $\chi'(\mathbf{q},\omega)$ around $\mathbf{q} = \mathbf{Q} = (\pi,\pi)$ and $\omega(\mathbf{Q}) = \Omega_{res}$. This 'across the centre of the zone' physics has to be contrasted with our 'to the corner of the zone' physics. To extract the large desired 'spin-fermion' coupling constant, $g$, and fermion self-energy coupling constant, $\lambda$, of roughly ¾ and 2 respectively in atomic units was shown to require that the 'magnetic' coherence length be just 2. Hence the bosonic coupling mode looked to within [45b] is of very highly overdamped spin excitations, and note contrary in this to what popularly is often implicated – an antiferromagnetic spin fluctuation mode acting as boson mediator in a BCS-related manner. However this $\xi$ value for the spin coherence length of just 2 means it is identical to the range of the superconductive coupling in the local pair view [2b,101], and it exposes the magnetic attribution of the coupling as not being unique. Certainly $g$ and $\lambda$ have to be large in order to yield HTSC, but the neutron peak has rather little spectral weight and role in events. Indeed the 'hot spots' on the Fermi surface dominate scattering activity in the zone and confer the novel transport behaviour above $T_c$ and the $d$-wave form to the superconductive gapping, but patently this is not due to Fermi surface nesting. The uncondensed boson modes dispersing upwards from the hot spots, that the STM scattering experiments from Davis and coworkers [1,63] would appear now to bear witness to, reaffirm the local pair character to events current in these materials, as of course above all does the extreme smallness of $\xi_o^{sc}$.

It seems to the author essential to try now to draw together the many disparate theoretical approaches followed in the past, including the spin fluctuation and MFL ones, around the negative-*U* and boson-fermion modelling being developed in some detail by de Llano, Tolmachev and colleagues [5,102], Tchernyshyov and Ren [103], Letz and Gooding [104], Domanski, Ranninger, Romani, Tripodi and coworkers [4,89,105], and various others. In doing this it will be very necessary to accommodate properly the inhomogeneous two-subsystem nature of the mixed-valent cuprates that the STM results in references [1,63] and related papers now so clearly affirm.

**Acknowledgments:** I would like to dedicate this paper to Prof B L Gyorffy on the occasion of his 65th birthday and for helping to make my time in Bristol a long, lively and productive one. At this moment of my own retirement I thank too my wife (P.A.W.) for her inestimable help over an even longer period in seeing this paper and its predecessors brought to fruition. Thanks are due too to Dr N E Hussey for his comments on the present manuscript.

.

**References**

[1]   Hoffman J E, McElroy K, Lee D-H, Lang K M, Eisaki H, Uchida S and Davis J C,
         2002 *Science* **297** 1148.                                                    [Jub 1184]

[2] a  Wilson J A,   2001 *J. Phys.: Condens. Matter* **13** R945-R977.
    b  Wilson J A,   2000 *J. Phys.: Condens. Matter* **12** R517-R547.
    c  Wilson J A,   2000 *J. Phys.: Condens. Matter* **12** 303-310.
    d  Wilson J A,   1998 *J. Phys.: Condens. Matter* **10** 3387-3410.
    e  Wilson J A and Zahrir A,   1997 *Rep. Prog. Phys.* **60** 941-1024.
    f  Wilson J A,   1994 *Physica* C **233** 332-348.
    g  Wilson J A,   1989 *Int. J. Mod. Phys.* B**3** 691-710.
    h  Wilson J A,   1988 *J. Phys. C; Solid State Phys.* **21** 2067-2102; 1987 ibid **20** L911-L916.

[3]   Micnas R,   2002 *arXiv:cond-mat*/0211561.                                        [Jû 2553]
      Micnas R, Robaszkiewicz S and Bussmann-Holder A,
         2002 *Phys. Rev.* B **66** 104516.                                               [Jû 2497]
      Micnas R and Tobijaszewska B,   2002 *J. Phys.: Condens. Matter* **14** 9631.     [Jû 2495]

[4]   Domanski T,   2002 *Phys.Rev.* B **66** 134512.                                   [Jû 2492]
      Domanski T and Wysokinski K I,   2002 *Phys. Rev.* B **66** 064517.              [Jû 2480]
      Domanski T and Ranninger J,   2002 *arXiv:cond-mat*/0208255.                      [Jû 2498]
      Domanski T, Maska M M and Mierzejewski M,   2002 *arXiv:cond-mat*/0207031.       [Jû 2485]

[5]   Batle J, Casas M, Fortes M, de Llano M, Rojo O, Sevilla F J, Solis M A
         and Tolmachev V V,   2002 *arXiv:cond-mat*/0211456.                            [Jû 2554]
      Casas M, de Llano M, Puente A, Rigo A and Solis M A,
         2002 *Solid State Commun.* **123** 101.                                         [Jû 2501]
      Fortes M, Solis M A, de Llano M and Tolmachev V V,   2001 *Physica* C **364/5** 95.  [Jû 2502]
      Batle J, Casas M, Fortes M, Solis M A, de Llano M, Valladares A A and Rojo O,
         2001 *Physica* C **364-365** 161.                                               [Jû 2204]

[6]   Loram J W, Mirza K A, and Cooper J R,   pp. 77-97 in *Research Review 1998 HTSC*.
         [Ed: Liang W Y;  Pub: IRC, Univ. of Cambridge, 1998].                          [Jup[2] 650]
      Loram J W, Luo J, Cooper J R, Liang W Y and Tallon J L,   2000 *Physica* C**341-8** 831,
         2001 *J. Phys. Chem. Solids* **62** 59.                                [Jub 1054],[Jup[2] 731]

[7]   Uemura Y J,   2000 *Int. J. Mod. Phys.* B **14** 3703.                            [Jup[4] 566]

[8]   Eschrig M and Norman M R,   2002 *arXiv:cond-mat*/0202083.                        [Jû 2445]
      Norman M,   2001 *Phys. Rev.* B **63** 092509.                                    [Jup[4] 562]
      Timm C, Manske D and Bennemann K H,   2002 *Phys. Rev.* B **66** 094515 + refs.  [Jû 2496]
      Yanase Y,   2002 *J. Phys. Soc. Jpn.* **71** 278.                                 [Jû 2512]
      Abanov A, Chubukov A V and Schmalian J,
         2001 *J. Electron Spectroscopy & related Phenomena* **117-118** 129.          [Jû 2386]

[9]   Fong H F, Bourges P, Sidis Y, Regnault L P, Ivanov A, Gu G D, Koshizuka N

and Keimer B,   1999 *Nature* **398** 588.                  [BSCCO-2212]   [Jub 892]

He H, Bourges P, Sidis Y, Ulrich C, Regnault L P, Pailhès S, Berzigiarova N S,
    Kolesnikov N N and Keimer B,   2002 *Science* **295** 1045.   [Tl-2201]   [Jut 424]

Dai P, Mook H A, Hayden S M, Aeppli G, Perring T G, Hunt R D and Dogan F,
    1999 *Science* **284** 1344.                  [YBCO-123]   [Jup$^4$ 534]

Keimer B, Aksay I A, Bossy J, Bourges P, Fong H F, Milius D L, Regnault L P
    and Vettier C,   1998 *J. Phys. Chem Solids* **59** 2135.   [YBCO-123]   [Jup$^4$ 541]

Fong H F, Bourges P, Sidis Y, Regnault L P, Bossy J, Ivanov A, Milius D L, Aksay I A
    and Keimer B,   2000 *Phys. Rev.* B **61** 14773.   [underdoped YBCO]   [Jup$^4$ 556]

[10]   Bogdanov P V, Lanzara A, Zhou X J, Kellar S A, Feng D L, Lu E D, Eisaki H, Shimoyama
        J-I, Kishio K, Hussain Z and Shen Z-X,   2001 *Phys. Rev.* B **64** 180505.   [Jub 1107]

       Feng D L, Kim C, Eisaki H, Lu D H, Damascelli A, Shen K M, Ronning F, Armitage N P,
           Kaneko N, Greven M, Shimoyama J-i, Kishio K, Yoshizaki R, Gu G D and Shen Z-X,
           2002 *Phys. Rev.* B **65** 220501(R).                [Jub 1059]

[11]   Chuang Y-D, Gromko A D, Federov A, Dessau A S, Aiura Y, Oka K, Ando Y, Eisaki H,
           Uchida S I and Dessau D S,   2001 *Phys. Rev. Lett.* **87** 117002.   [Jub 1042]

       Chuang Y-D, Gromko A D, Fedorov A, Aiura Y, Oka K, Ando Y and Dessau D S,
           2001 *arXiv:cond-mat*/0107002.                       '[Jub 1145]'

[12]   Kordyuk A A, Borisenko S V, Kim T K, Nenkov K A, Knupfer M, Fink J, Golden M S,
           Berger H and Follath R,   2002 *Phys. Rev. Lett.* **89** 077003.   [Jub 1194]

       Borisenko S V, Kordyuk A A, Kim T K, Legner S, Nenkov K A, Knupfer M, Golden M S,
           Fink J, Berger H and Follath R,   2002 *arXiv:cond-mat*/0204557.   [Jub 1177]

       Borisenko S V, Kordyuk A A, Kim T K, Koitzsch A, Knupfer M, Golden M S, Fink J,
           Eschrig M, Berger H and Follath R,   2002 *arXiv:cond-mat*/0209435.   [Jub 1188]

[13]   Feng D L, Damascelli A, Shen K M, Motoyama N, Lu D H, Eisaki H, Shimizu K,
           Shimoyama J-i, Kishio K, Kaneko N, Greven M, Gu G D, Zhou X J, Kim C, Ronning F,
           Armitage N P and Shen Z-X,   2002 *Phys. Rev. Lett.* **88** 107001.   [Jub 1062]

[14]   Sato T, Matsui H, Nishina S, Takahashi T, Fujii T, Watanabe T and Matsuda A,
           2002 *Phys. Rev. Lett.* **89** 067005.                [Jub 1190]

       Takeuchi T, Yokoya T, Shin S, Jinno K, Matsuura M, Kondo T, Ikuta H and Mizutani U,
           2001 *J. Elec Spectr. & Rel. Phen.* **114/6** 629.       [Jub >]

[15]   Campuzano J C, Norman M R and Randeria M,   2002 *arXiv:cond-mat*/0209476. [Jup$^3$ 554]

[16]   Anderson O K, Jepsen O, Liechtenstein A I, and Mazin I I,
           1994 *Phys. Rev.* B **49** 4145.                      [Jû 1121]

       Massida S, Yu J and Freeman A J,   1988 *Physica* C **152** 251.   [Jû 378]

[17]   Meeson P J,   private communication.

[18]   Goodrich R G, Grenier C, Hall D, Lacerda A, Haanappel E G, Rickel D, Northington T,
           Schwarz R, Mueller F M, Koelling D D, Vuillemin J, van Bockstal L, Norton M L
           and Lowndes D H,   1993 *J. Phys. Chem. Solids* **54** 1251.   [Jπ 268]

N.B. This high field work was at the limit of detection, and figures 5 and 6 seem not compatible.

[19]   Hussey N E, McBrien M N, Balicas L, Brooks J S, Horii S and Ikuta H,
      2002 *Phys. Rev. Lett.* **89** 086601.     [Juf 490]

McBrien M N, Hussey N E, Meeson P J, Horii S and Ikuta H,
      2002 *J. Phys. Soc. Jpn.* **71** 701.     [Juf 489]

Horii S, Takagi H, Ikuta H, Hussey N E, Hirabayashi I and Mizutani U,
      2002 *Phys. Rev.* B **66** 054530.     [Juf 491]

[20]   Wilson J A and Farbod M,   2000 *Supercond. Sci. Technol.* **13** 307.
N.B. The way to obtain fully Hg-loaded Hg-1201 samples is now known:
    Alyoshin V A, Mikhailova D A, Rudnyi E B and Antipov E V,   2002 *Physica* C **383** 59. [Juh 295]

[21]   Hussey N E,   2003 *Eur. Phys. J* B **31** 495.     [Jû 2499]

[22]   Wilson J A,   1972 *Adv. in Phys.* **21** 143-198.

[23]   Takagi H, Batlogg B, Kao H L, Kwo J, Cava R J, Krajewski J J, and Peck W F,
      1992 *Phys. Rev. Lett.* **69** 2975.     [Jul 607]

Kubo Y and Manako T,   1992 *Physica* C **197** 388.     [Jut 251]
                         1994 *Phys. Rev.* B **50** 6402.     [Jut 303]

Nakamae S, Behnia K, Mangkorntong N, Nohara M, Takagi H, Yates S J C
   and Hussey N E,   2002 *arXiv:cond-mat*/0212283.     [Jul 1127]

[24]   Mesot J, Norman M R, Ding H, Randeria M, Campuzano J C, Paramekanti A, Fretwell H M,
   Kaminski A, Takeuchi T, Yokoya T, Sato T, Takahashi T, Mochiku T and
     Kadowaki K,   1999 *Phys. Rev. Lett.* **83** 840;     [Jub 950]
                           1999 *J. Low Temp. Phys.* **117** 365.     [Jub 961]

[25]   Norman M R and Ding H,   1998 *Phys. Rev.* B **57** 11089.     [Jub 871]

Norman MR, Randeria M, Ding H and Campuzano J C,
   1998 *Phys. Rev.* B **57** R11093.     [Jub 872]

Norman M R, Kaminski A, Mesot J and Campuzano J C,
   2001 *Phys. Rev.* B **63** 140508(R).     [Jub 1050]

[26]   Johnson P D, Valla T, Federov A V, Yusof Z, Wells B O, Li Q, Moodenbaugh A R,
  Gu G D, Koshizuka N, Kendziora C, Jian S, Hinks D G,
   2001 *Phys. Rev. Lett.* **87** 177007.     [Jub 1037]

[27]   Lanzara A, Bogdanov P V, Zhou X J, Kellar S A, Feng D L, Lu E D, Yoshida T,
   Eisaki H, Fujimori A, Kishio K, Shimoyama J, Noda T, Uchida S, Hussain Z
     and Shen Z-X,   2001 *Nature* **412** 510.     [Jub 1038]

[28]   Valla T, Federov A V, Johnson P D and Hulbert S L,   1999 *Phys. Rev. Lett.* **83** 2085.   [*]

[29]   Valla T, Federov A V, Johnson P D, Xue J, Smith K E and DiSalvo F J,
   2000 *Phys. Rev. Lett.* **85** 4759.     [Cε' 93]

[30]   Eschrig M and Norman M R,   2000 *Phys. Rev. Lett.* **85** 3261.     [Jû 2135]

Eschrig M and Norman M R,   2002 *Phys. Rev. Lett.* **89** 277005.     [Jû 2500]

[31]   Stevens C J, Smith D, Chen C, Ryan J F, Pobodnik B, Mihailovic D, Wagner G A and

| | | |
|---|---|---|
| | Evetts J E, 1997 *Phys. Rev. Lett.* **78** 2212. | [Jup³ 450] |
| [32] | Holcomb M J, Perry C I, Collman J P and Little W A, 1996 *Phys. Rev.* B **53** 6734. | [Jup³ 436] |
| [33] | Little W A and Holcomb M J, 2000 *J. Supercond.* **13** 695. | [Jû 2189] |
| | Little W A, Collins K and Holcomb M J, 1999 *J. Supercond.* **12** 89. | [Jû 2206] |
| [34] | Kabanov V V, Demsar J, Pobodnik B and Mihailovic D, 1999 *Phys. Rev.* B **59** 1497. | [p³ 485] |
| | Demsar J, Podobnik B, Kabanov V V, Wolf Th and Mihailovic D, 1999 *Phys. Rev. Lett.* **82** 4918. | [Jup³ 473] |
| | Kabanov V V, Demsar J and Mihailovic D, 2000 *Phys. Rev.* B **61** 1477. | [Jup³ 528] |
| | Demsar J, Hudej R, Karpinski J, Kabanov V V and Mihailovic D, 2001 *Phys. Rev.* B **63** 054519. | [Juh 274] |
| [35] | Li E, Li J J, Sharma R P, Ogale S B, Cao W L, Zhao Y G, Lee C H and Venkatesan T, 2002 *Phys. Rev.* B **65** 184519. | [Jup³ 546] |
| [36] | Gyorffy B L, Staunton J B and Stocks G M, 1991 *Phys. Rev.* B **44** 5190. | [Jû 855] |
| | Litak G and Gyorffy B L, 2000 *Phys. Rev.* B **62** 6629. | [Jû 2120] |
| | Kopeć T K, 2002 *Phys. Rev.* B **66** 184504. | [Jû 2550] |
| | Chen Q, Levin K and Kosztin I, 2001 *Phys. Rev.* B **63** 184519, & refs. therein. | [Jû 2213] |
| [37] | Micnas R, Ranninger J and Robaszkiewicz S, 1990 *Rev. Mod. Phys.* **62** 113. | [Jû 596] |
| [38] | Friedberg R and Lee T D, 1989 *Phys. Rev.* B **40** 6745. | [Jû 517] |
| [39] | Corson J, Mallozi R, Orenstein J, Eckstein J N and Bozovic I, 1999 *Nature* **398** 221. | [b 888] |
| | Corson J, Orenstein J, Oh S, O'Donnell J and Eckstein J N, 2000 *Phys. Rev. Lett.* **85** 2569. | [Jub 1022] |
| [40] | Bourges P, Casalta H, Regnault L P, Bossy J, Burlet P, Vettier C, Beaugnon E, Gautier-Picard P and Tournier R, 1997 *Physica* B **234/6** 830. | [Jup⁴ 570] |
| | Mook H A, Dai P, Hayden S M, Hiess A, Lynn J W, Lee S-H and Dogan F, 2002 *arXiv:cond-mat*/0204002. | [Jup⁴ 592] |
| | Zavidonov A Yu and Brinkmann D, 2001 *Phys. Rev.* B **63** 132506. | [Jup⁴ 594] |
| [41] | Reichardt W, 1996 *J. Low Temp. Phys.* **105** 807. | [Ju ***] |
| | Pintschovius L, Reichardt W, Kläser M, Wolf T and v. Löhneysen H, 2002 *Phys. Rev. Lett.* **89** 037001. | [Jup⁴ 586] |
| | McQueeney R J, Sarrao J L, Pagliuso P G, Stephens P W and Osborn R, 2001 *Phys. Rev. Lett.* **87** 077001. | [Jul 1027] |
| [42] | Tajima S, Schützmann J, Miyamoto S, Terasaki I, Sato Y and Hauff R, 1997 *Phys. Rev.* B **55** 6051. | [Jup³ 472] |
| | Ioffe L B and Millis A J, 2000 *Phys. Rev.* B **61** 9077. | [Jû 2150] |
| [43] | Krishana K, Harris J M and Ong N P, 1995 *Phys. Rev. Lett.* **75** 3529. | [Jup² 588] |
| [44] | Takigawa M and Mitzi D B, 1994 *Phys. Rev. Lett.* **73** 1287. | [Jub 682] |
| [45] | Kee H-Y, Kivelson S A and Aeppli G, 2002 *Phys. Rev. Lett.* **88** 257002. | [Jû 2416] |
| | Abanov A R, Chubukov A V, Eschrig M, Norman M R and Schmalian J, | |

|  | 2002 *Phys. Rev.* B **89** 177002. | [Jû 2433] |

[46] Reznik D, Pintschovius L, Reichardt W, Endoh Y, Hiraka H, Tranquada J M, Tajima S,
     Uchiyama S and Masui T,   2002 *arXiv:cond-mat/*0209197.   [Jup$^4$ >]

[47] Moncton D E, Axe J D and DiSalvo F J,   1977 *Phys. Rev.* B **16** 801. [2H-TaSe2]   [Cε 96]

[48] Belkhir L and Randeria M,   1994 *Phys.Rev* B **49** 6829.   [Jû 1115]

[49] Li Y, Huang J L and Lieber C M,   1992 *Phys. Rev. Lett.* **68** 3240.   [Jub 550]
     Li Y and Lieber C M,   1993 *Mod. Phys. Lett.* B **7** 143.   [Jup$^3$ 363]

[50] Wang Y, Ono S, Onose Y, Gu G, Ando Y, Tokura Y, Uchida S and Ong N P,
     2003 *Science* **299** 86.   [Jub 1219]

[51] Loram J W, Tallon J L and Liang W Y,   2002 *arXiv:cond-mat/*0212461.   [Jup$^2$ 784]

[52] Pan S H, O'Neal J P, Badzey R L, Chamon C, Ding H, Engelbrecht J R, Wang Z, Eisaki H,
     Uchida S, Gupta A K, Ng K-W, Hudson E W, Lang K M and Davis J C,
     2001 *Nature* **413** 282.   [Jub 1057]

[53] Singer P M, Hunt A W, Cederström A F and Imai T,
     2003 *arXiv:cond-mat/*0302077.   [Jul 1128]
     Singer P M, Hunt A W and Imai T,   2003 *arXiv:cond-mat/*0302078.   [Jul 1129]

[54] Farbod M, Giblin S, Bennett M and Wilson J A,   2000 *J. Phys.: Condens. Matter* **12** 2043.
     Wilson J A and Farbod M,   2000 *Supercond. Sci. Technol.* **13** 307.

[55] Watanabe I, Kawano K, Kumagai K, Nishiyama K and Nagamine K,
     1992 *J. Phys. Soc. Jpn.* **61** 3058.   [Jul 555]
     Note the results of -
     Sonier J E, Brewer J H, Kiefl R F, Heffner R H, Poon K F, Stubbs S L, Morris G D,
     Miller R I, Hardy W N, Liang R, Bonn D A, Gardner J S & Stronach C E and Curro N J,
     2003 *Phys. Rev.* B **66** 134501,   [Jup$^4$ 597]
     refute their earlier conclusions published in -
     Sonier J E, Brewer J H, Kiefl R F, Miller R I, Morris G D, Stronach C E, Gardner J S,
     Dunsiger S R, Bonn D A, Hardy W N, Liang R and Heffner R H,
     2001 *Science* **292** 1692.   [Jup$^4$ 567]

[56] Mason T E, Aeppli G, Hayden S M, Ramirez A P and Mook H A,
     1993 *Phys. Rev. Lett.* **71** 919.   [Jul 567]

[57] Radcliffe J W, Loram J W, Wade J M, Witschek G and Tallon J L,
     1996 *J Low Temp. Phys.* **105** 903.   [Jut 369]

[58] Muller K A, Takashige M and Bednorz J G,   1987 *Phys. Rev. Lett.* **58** 1143.   [Ju 143]
     Deutscher G and Müller K A,   1987 *Phys. Rev. Lett.* **59** 1745.   [Jû 121]

[59] Lang K M, Madhavan V, Hoffman J E, Hudson E W, Eisaki H, Uchida S and Davis J C,
     2002 *Nature* **415** 412.   [Jub 1159]

[60] Suvasini M B, Temmerman W M and Gyorffy B L,   1993 *Phys. Rev.* B **48** 1202.   [Jû *****]

[61] Ghosal A, Randeria M and Trivedi N,   2001 *Phys. Rev.* B **65** 014501.   [Jû 2389]

[62] Wang Z, Engelbrecht J R, Wang S, Ding H and Pan S H,
     2002 *Phys. Rev.* B **65** 064509.   [Jû 2579]

[63] McElroy K, Simmonds R W, Hoffmann J E, Lee D-H, Orenstein J, Eisaki H, Uchida S
 and Davis J C,  2003 *Nature* **********.  [Jub 1209]

[64] Chung J-H, Egami T, McQueeney R J, Yethiraj M, Arai M, Yokoo T, Petrov Y, Mook H M,
 Endoh Y, Tajima S, Frost C and Dogan F,  2003 *Phys. Rev.* B **67** 014517. [Jup[4] 595]

[65] Chubukov A V, Jankó B and Tchernyshyov O,  2001 *Phys. Rev.* B **63** 180507(R).[Jû 2171]

[66] Mook H A, Dai P, Dogan F and Hunt R D,  2000 *Nature* **404** 729.  [Jup[4] 554]

[67] Munzar D and Cardona M,  2003 *Phys. Rev. Lett.* **90** 077001.  [Jub1220]

[68] Cardona M,  1999 *Physica* C **317/8** 30-54.  [Jup[3] 508]

[69] Munzar D, Bernhard C, Holden T, Golnik A, Humlicek J and Cardona M,
 2001 *Phys. Rev.* B **64** 024523.  [Jup[3] 551]

 Timusk T,  2003 *arXiv:cond-mat*/0303383.  [Jup[3] 581]

[70] McGuire J J, Windt M, Startseva T, Timusk T, Colson D, Viallet-Guillen V,
 2000 *Phys. Rev.* B **62** 8711.  [Juh 271]

[71] Keller H,  2003 *Physica* B **326** 283.  [Jû 2582]

[72] Matsui H, SatoT, Takahashi T, Ding H, Yang H-B, Wang S-C, Fujii T, Watanabe T,
 Matsuda A, Terashima T and Kadowaki K,  2003 *Phys. Rev.* B **64** 060501(R). [Jub 1222]

[73] Schilling J S,  2001 *arXiv:cond-mat*/0110267.  [Jup[2] 771]

[74] Kim T K, Kordyuk A A, Borisenko S V, Koitzsch A, Knupfer M, Berger H and Fink J,
 *arXiv:cond-mat*/0303422.  [Jub 1224]

[75] Domanski T,  2003 *arXiv:cond-mat*/0302406.  [Jû 2581]

[76] Hirsch J E,  2003 *Phys. Rev.* B **67** 035103.  [Jû 2552]

[77] Eremets M I, Struzkhin VV, Mao H-K and Hemley R J,  2001 *Science* **293** 272.  [Lv  ]

[78] Struzhkin V V, Eremets M I, Gan W, Mao H-K and Hemley R J,
 2002 *Science* **298** 1213.  [Lv  ]

 Shimitsu K, Ishikawa H, Takao D, Yagi T and Amaya K,  2002 *Nature* **419** 597.  [Lv  ]

[79] Wilson J A,  1991 *Physica* C **182** 1.

[80] Hagel J, Kelemen M T, Fischer G, Pilawa B, Wosnitsa J, Dormann E, v. Lohneysen H,
 Schnepf A, Schockel H, Neisel U and Beck J,
 2002 *J. Low Temp. Phys.* **129** 133.  [Lv  ]

[81] Ricardo da Silva R, Torres J H S and Kopelevich Y,  2001 *Phys. Rev. Lett.* **87** 147001.[v   ]

[82] Boeri L, Bachelet G B, Cappelluti E and Pietronero L,
 2003 *Supercond. Sci Tech.* **16** 143.  [Lβ 246]

[83] Mazin I I and Antropov V P,  2003 *Physica* C **385** 49.  [Lβ 247]

[84] Wilson J A,  1999 *Supercond. Sci. Technol.* **12** 649-653.

[85] Wen H-H, Yang H-P, Lu X-F and Yan Y,  2003 *Chin. Phys. Lett.* **20** 725.  [Jρ 82]

 Takada K, Sakurai H, Takayama-Muromachi E, Izumi F, Dilanian R A and Sasaki T,

|       | 2003 *Nature* **422** 53. | [Jρ 83] |
| [86]  | Alexandrov A S, 2003 *arXiv:cond-mat*/0303608. | [Jû 2587] |
| [87]  | Tan S and Levin K, 2003 *arXiv:cond-mat*/0302248. | [Jû 2584] |
| [88]  | Emery V J, Kivelson S A and Zacher O, 1997 *Phys. Rev.* B **56** 6120. | [Jû 1714] |
| [89]  | Ranninger J and Tripodi L, 2002 *arXiv:cond-mat*/0212332. | [Jû 2555] |
| [90]  | Nandor V A, Martindale J A, Groves R W, Vyaselev O M, Pennington C H, Hults L and Smith J L, 1999 *Phys. Rev.* B **60** 6907. | [Jup[4] 540] |
|       | Mitrovic' V F, Bachmann H N, Halperin W P, Reyes A P, Kuhns P and Moulton W G, 2001 *Nature* **413** 501. | [Jup[4] 568] |
| [91]  | Kordyuk A A, Borisenko S V, Knupfer M and Fink J, 2003 *Phys. Rev.* B **67** 064504. | [Jub 1221] |
| [92]  | Verga S, Knigavko A and Marsiglio F, 2003 *Phys. Rev.* B **67** 054503. | [Jul 1130] |
| [93]  | Lake B, Aeppli G, Mason T E, Schröder A, McMorrow D F, Lefmann K, Isshiki M, Nohara M, Takagi H and Hayden S M, 1999 *Nature* **400** 43. | [Jub 927] |
| [94]  | Millis A J and Drew H D, 2003 *arXiv:cond-mat*/0303018. | [Jû 2583] |
| [95]  | Abrahams E and Varma C M, 2000 *arXiv:cond-mat*/0003135. | [Jû 2178] |
|       | Abrahams E and Varma C M, 2000 *Proc. Natl. Acad. Sci.* **97** 5714. | [Jû    ] |
| [96]  | Timusk T and Statt B, 1999 *Rep. Prog. Phys.* **62** 61. | [Jup[3] 516] |
| [97]  | Varma C M, 1989 *Int. J. Mod. Phys.* B **3** 2083. | [Jû 578] |
| [98]  | Varma C M and Abrahams E, 2001 *Phys. Rev. Lett.* **86** 4652. | [Jû 2285] |
| [99]  | Varma C M and Abrahams E, 2002 *arXiv:cond-mat*/0202040. | [Jû 2285b] |
|       | Carter E C and Schofield A J, 2002 *arXiv:cond-mat*/0209003. | [Jû 2542] |
| [100] | Abanov A R, Chubukov A V, Eschrig M, Norman M R and Schmalian J, | |
| ,     | 2001 *arXiv:cond-mat*/0112126 | [Jû 2433] |
| [101] | Quintanilla J and Gyorffy B L, 2002 *J. Phys.: Condens. Matter* B **14** 6591. | [Jû 2222] |
| [102] | de Llano M and Tolmachev VV, 2003 *Physica* A **317** 546. | [Jû 2577] |
| [103] | Tchernyshyov O, 1997 *Phys. Rev.* B **56** 3372. | [Jû 2166] |
|       | Ren H-C, unpublished,: see 1996 *Bull. Amer. Phys. Soc.* **41** 183 (E13.7;13.9). | [Jû 2163] |
| [104] | Letz M, 1999 *J. Supercond.* **12** 61, and refs. therein. | [Jû 2207] |
| [105] | Domanski T and Ranninger J, 2001 *Phys. Rev.* B **63** 134505. | [Jû 2202] |
|       | Ranninger J and Romano A, 2002 *arXiv:cond-mat*/0207189. | [Jû 2490] |